\documentclass[aps,prb,showpacs,superscriptaddress]{revtex4-1}
\usepackage{times}
\usepackage{epsfig}
\usepackage{graphicx}
\usepackage{amsmath}
\begin{document}
\title{Long-lived selective spin echoes in dipolar solids under periodic and aperiodic $\pi-$pulse trains}
\author{Clark D. Ridge}
\thanks{Current Address:  FDA, College Park, Maryland 20740, USA}
\affiliation{Department of Chemistry,
 University of Miami, Coral Gables, Florida 33124,USA}
         \author{Lauren F. O'Donnell}
        \affiliation{Department of Chemistry,
         University of Miami, Coral Gables, Florida 33124,USA}
\author{Jamie D. Walls}
\email{Corresponding author: jwalls@miami.edu}
\affiliation{Department of Chemistry,
 University of Miami, Coral Gables, Florida 33124,USA}
\date{\today}
\begin{abstract}
 The application of  Carr-Purcell-Meiboom-Gill (CPMG) $\pi-$trains for dynamically decoupling a system from its environment has been extensively studied in a variety of physical systems.  When applied to dipolar solids, recent experiments have demonstrated that CPMG pulse trains can generate long-lived spin echoes.  While there still remains some controversy as to the origins of these long-lived spin echoes under the CPMG sequence, there is a general agreement that pulse errors during the $\pi-$pulses are a necessary requirement.  In this work, we develop a theory to describe the spin dynamics in dipolar coupled spin-1/2 system under a CPMG($\phi_{1},\phi_{2}$) pulse train, where $\phi_{1}$ and $\phi_{2}$ are the phases of the $\pi-$pulses.  From our theoretical framework, the propagator for the CPMG($\phi_{1},\phi_{2}$) pulse train is equivalent to an effective ``pulsed'' spin-locking of single-quantum coherences with phase $\pm\frac{\phi_{2}-3\phi_{1}}{2}$, which generates a periodic quasiequilibrium that corresponds to the long-lived echoes.  Numerical simulations, along with experiments on both magnetically dilute, random spin networks found in C$_{60}$ and C$_{70}$ and in non-dilute spin systems found in adamantane and ferrocene, were performed and confirm the predictions from the proposed theory.
\end{abstract}
\maketitle
\section{Introduction}
\begin{figure}
\includegraphics*[scale=.4]{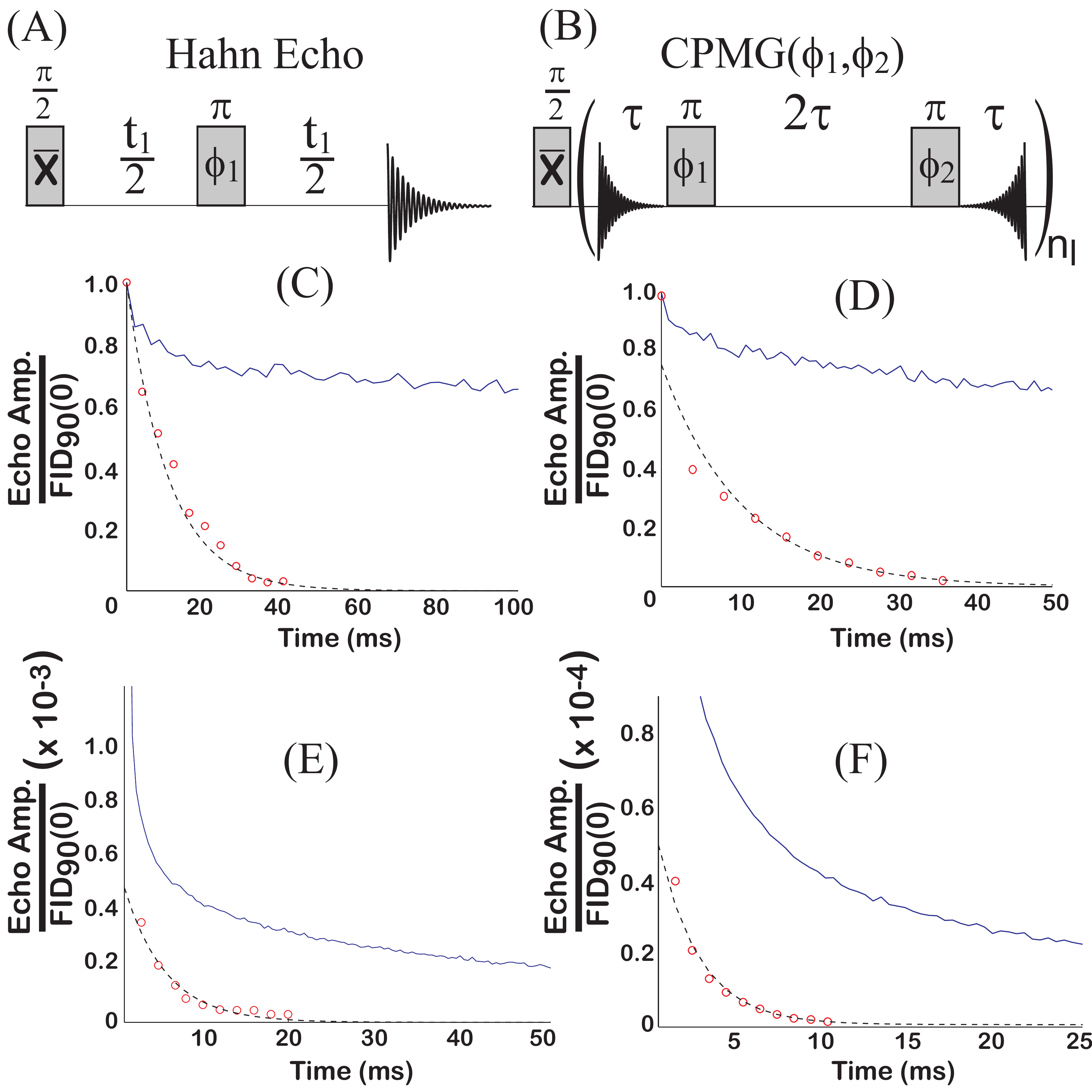}
\caption{(Color online) (A) Hahn echo (single $\pi-$pulse) and (B) CPMG($\phi_{1},\phi_{2}$) experiments.  In a Hahn echo experiment, the echo amplitudes are collected from multiple measurements using different values of $t_{1}$, whereas in the CPMG($\phi_{1},\phi_{2}$) experiment, echo amplitudes at integer multiples of $4\tau$ are collected within a single measurement.  In Figs. \ref{fig:fig1}(C)-\ref{fig:fig1}(F), the Hahn (circular red dots) and CPMG($Y,\,Y$) echo amplitudes (blue curve, sampled every $4\tau$) are shown for (C) $C_{60}$ [N$_s$ (Number of scans) = 32 for the Hahn echoes;  N$_s$=128 and $\tau=500\mu$s for the CPMG($Y,\,Y$) echoes], (D) $C_{70}$ [N$_s$=256 for the Hahn echoes; N$_s$=512 and $\tau=200\mu$s for the CPMG($Y,\,Y$) echoes], (E) adamantane (C$_{10}$H$_{16}$) [N$_s$=128 for the Hahn echoes; N$_s$=128 and $\tau=116\mu$s for the CPMG($Y,\,Y$) echoes], and (F) ferrocene (Fe$\left(\text{C}_{5}\text{H}_{5}\right)_{2}$) [N$_s$=128 for the Hahn echo experiment; N$_s$=256 and $\tau=116\mu$s for the  CPMG($Y,\,Y$) echoes].  The echo amplitudes were normalized by the signal from a $\frac{\pi}{2}-$acquire experiment with an equivalent number of scans.  In all cases, the Hahn echo amplitudes decayed at a faster rate than the corresponding CPMG($Y,\,Y$) echo amplitudes.  Exponential fits of the Hahn echo data $\left[Aexp\left(-\Gamma t\right)\right]$ are shown (black dotted curve) with decay constants of (C) $\Gamma=94 s^{-1}$, (D) $\Gamma=97 s^{-1}$, (E) $\Gamma=204 s^{-1}$, and (F) $\Gamma=410 s^{-1}$.
}
\label{fig:fig1}
\end{figure}

Preserving quantum coherence is of importance to a variety of fields, from quantum computing and information processing\cite{Chirolli08, Hanson07}, where unitary control of qubits is a critical requirement of many proposed algorithms, to biomedical applications, where long-lived signals can help improve resolution in imaging applications\cite{Frey12,zhang13}.  As such, a variety of techniques have been developed to increase the lifetime of the coherence being studied by controlling the system-environment interactions. In dynamical decoupling\cite{Viola98,Uhrig09}, a series of control fields are applied to the system so that, on average, the system is effectively decoupled from its environment.  One common method of dynamical decoupling is the application of inversion or $\pi-$pulses to the system that are capable of preserving coherence created by an initial excitation.  Examples of such dynamical decoupling sequences are the [Fig. \ref{fig:fig1}(A)] Hahn or spin echo\cite{Hahn50} and the [Fig. \ref{fig:fig1}(B)] Carr-Purcell-Meiboom-Gill (CPMG) sequence\cite{Carr54}.  These sequences refocus system-environment interactions, $\widehat{H}_{SE}$, that are odd with respect to inversion by a $\pi-$pulse applied to the system, e.g., the heteronuclear dipolar interaction between an $I$ spin (i.e., the system) and a collection of $S$ spins (i.e., the environment)\cite{Alvarez10a}.    The efficacy of these sequences in preserving quantum coherence, even in the presence of $\pi-$pulse errors, has been studied extensively in a variety of systems\cite{Alvarez10a,Souza11,Wang12,Wang12a,Bar-Gill12}.

 While the $\pi-$pulses can refocus $\widehat{H}_{SE}$ if it is odd under inversion by a $\pi-$pulse applied to the system, the system's Hamiltonian, $\widehat{H}_{S}$, can also be affected by the dynamical decoupling sequence.  This has been shown to be the case in dipolar solids, a potential platform for quantum computing/information processing. In solids, the main spin interactions for a system of spin 1/2 nuclei are the homonuclear dipole-dipole Hamiltonian, $\widehat{H}_{D}$, anfrd the chemical shift interaction, $\widehat{H}_{cs}$.  In an ``ideal" Hahn echo experiment\cite{Hahn50} [Fig. \ref{fig:fig1}(A)], $\widehat{H}_{cs}$ is refocused on a time-scale of $t_{1}$ due to the application of a radiofrequency (RF) pulse of flip angle 180$^{\circ}$, referred to as a $\pi-$pulse, applied at a time $\frac{t_{1}}{2}$.  Due to the $\pi-$pulse, the system effectively evolves under $\widehat{H}_{D}$ for a time $t_{1}$.  However, additional corrections to the evolution on the order of $\left|\left[\widehat{H}_{D},\,\widehat{H}_{cs}\right]t_{1}\right|$ contribute at large $t_{1}$.  In order to reduce these higher-order corrections, a CPMG pulse train\cite{Carr54} can be used to refocus $\widehat{H}_{cs}$ on a faster time scale while still enabling the system to evolve under $\widehat{H}_{D}$ for long periods of time.  The CPMG($\phi_{1},\phi_{2}$) pulse train consists of a series of two $\pi-$pulses with phases $\phi_{1}$ and $\phi_{2}$ and separated by a time $2\tau$ that is repeated $n_{l}$ times [Fig. \ref{fig:fig1}(B)].  In Fig. \ref{fig:fig1}(B), even echoes occur at integer multiples of $4\tau$ [note that odd echoes also occur at $4n\tau+2\tau$ for $n=0,\,1,\,2,\,...$, but in this work, we focus on the even echoes since we are interested in studying the effects of $\pi-$pulse phases, $\phi_{1}$ and $\phi_{2}$].  In a CPMG experiment, $n_{l}$ echo amplitudes can be recorded in a single measurement, which is in contrast to the Hahn echo experiment, where $n_{l}$ different experiments are needed to record the echo amplitudes at times $t_{1}=4\tau$ to $t_{1}=4n_{l}\tau$.  If $\widehat{H}_{cs}$ is negligible or if $[\widehat{H}_{cs},\widehat{H}_{D}]=0$, then both the CPMG and the Hahn echo experiments should give identical echo f amplitudes as a function of time, which are given by the free induction decay (FID) under $\widehat{H}_{D}$, FID(t) = $\text{Trace}\left[ (\widehat{I}_{X}+i\widehat{I}_{Y})\widehat{\rho}(t)\right]$, where $\widehat{I}_{X}$ and $\widehat{I}_{Y}$ represent the total spin polarization along the $\widehat{x}$ and $\widehat{y}$ directions, respectively, and $\widehat{\rho}(t)$ is the density matrix at time $t$.   However, the echo amplitudes from the CPMG experiments were found\cite{Dementyev03,Watanabe03} to decay at a much slower rate than those found from the corresponding Hahn echo experiments.  Such observations are reconfirmed in Figs. \ref{fig:fig1}(C)-\ref{fig:fig1}(F), where the Hahn echoes [red, circular dots along with numerically fitted exponential decay curve (black dotted line)] are shown along with the echoes from a CPMG($Y,\,Y$) experiment in magnetically dilute $^{13}$C spin systems [Fig. \ref{fig:fig1}(C)] C$_{60}$ and [Fig. \ref{fig:fig1}(D)] C$_{70}$ and in non-dilute $^{1}$H systems [Fig. \ref{fig:fig1}(E)] adamantane and [Fig. \ref{fig:fig1}(F)] ferrocene.

There have been two main theoretical proposals to understand the long-lived spin echoes under CPMG pulse trains.  One explanation, propounded by the Barrett group\cite{Li07,Li08,Dong08}, which we refer to in throughout the paper as the Barrett proposal, is that these long-lived echoes are a consequence of resonance offset and dipolar evolution during the $\pi-$pulse, i.e., the $\pi-$pulses cannot be treated as pure $\delta$-pulses.  Using average Hamiltonian theory\cite{Haeberlen68} (AHT), these authors demonstrated that the CPMG($\phi_{1},\phi_{2}$) generates an effective field that, depending on the $\pi$ pulse phases, $\phi_{1}$ and $\phi_{2}$, can spin-lock the magnetization after an initial $\left(\frac{\pi}{2}\right)_{X}$ pulse; it was argued\cite{Li08} that this effect was not equivalent to pulsed spin-locking.  In their studies, the time $\tau$ in the CPMG pulse train was often on the same order as the $\pi-$pulse time, $t_{p}$.  However, it was also observed that long-lived echoes could also be observed even when $\tau\gg t_{p}$, where many of the AHT arguments used in the Barrett proposal\cite{Li07,Li08,Dong08} would break down since the contribution of $\widehat{H}_{D}$ during the pulse to the average Hamiltonian would appear to be negligible under these conditions.  An alternative theory proposed by the Levstein group\cite{Franzoni05,Franzoni08,Levstein08,Franzoni12}, which will be referred to as the Levstein proposal in this paper, was advanced which argued that the long-lived echoes were not due to violations of the $\delta$-pulse limit but were instead the consequence of imperfect $\pi-$rotations from field inhomogeneities within the sample along with the absence of spin diffusion during the CPMG sequence.  In their proposal, the imperfect $\pi-$pulses could store some of the single-quantum coherence along the the direction of the large, static Zeeman field (taken to be along the $\widehat{z}$-direction) in the form of $\widehat{z}$-magnetization and/or zero-quantum coherences.  $\widehat{z}$-magnetization or population tends to relax at a slower rate that is inversely proportional to the longitudinal relaxation time, T$_1$, relative to the relaxation decay of single-quantum coherences, which is inversely proportional to the transverse relaxation time, T$_2$.  In the absence of significant spin diffusion, subsequent imperfect $\pi-$pulses could restore the zero-quantum coherences/populations back into observable single-quantum coherence to form stimulated echoes\cite{Hahn50}.  Since it is often the case in solids that T$_1\gg$ T$_2$, the stimulated echo would be less attenuated by relaxation than the regular spin echo generated from signal that remained single-quantum coherence throughout the $\pi-$pulse train.  Levstein argued that the contributions from the stimulated echoes therefore make the echo amplitudes appear to be ``artificially'' long-lived.  Because it was demonstrated that stimulated echoes in C$_{60}$ were present after the application of a few $\pi-$pulses\cite{Franzoni08, Franzoni12}, it was argued that stimulated echoes generated by magnetization coherence transfer pathways contributed at all echo times due to the periodicity of the CPMG pulse train\cite{Goelman95,Hung10}.

  	In this work, we present an alternative theory for the long-lived echoes in dipolar solids observed under CPMG($\phi_{1},\phi_{2}$) pulse trains under the conditions $\tau\gg t_{p}$.   We demonstrate that the form of the propagator for a CPMG($\phi_{1},\phi_{2}$) pulse block with imperfect $\pi-$pulses (due to resonance offsets and RF flip-angle errors and neglecting $\widehat{H}_{D}$ during the $\pi$-pulses) is similar to the propagator for pulsed spin-locking in dipolar solids\cite{Ostroff66,Rhim76,Suwelack80,Ivanov78,Maricq85} with the phase of the effective ``spin-locking pulses'' given by $\frac{\phi_{2}-3\phi_{1}}{2}$ when $2(\phi_{2}-\phi_{1})=2n\pi$ for integer $n$ (it should also be noted that previous work\cite{Alvarez10a} has also suggested that spin-locking can be generated by a CPMG pulse train).  Under this effective ``pulsed'' spin-locking generated by the CPMG pulse train, a periodic quasiequilibrium\cite{Sakellariou98} that corresponds to the observed long-lived echoes under the CPMG($\phi_{1},\,\phi_{2}$) pulse train can be generated only if the initial phase of the magnetization, $\phi_{init}$, is not orthogonal to the phase of the effective spin-locking field, i.e.,  $\phi_{init}\neq\pm(\frac{\phi_{2}-3\phi_{1}}{2}+\frac{\pi}{2})$.  Unlike the Levstein and Barrett proposals, a reduction in the signal from the long-lived echoes along with a narrowing of the spectrum from the last echo generated by the CPMG($\phi_{1},\phi_2)$ pulse train are predicted to occur as the interpulse spacing ($2\tau$) increases.  In our calculations and simulations, relaxation effects were also neglected, but long-lived echoes were still observed without the requirement that T$_1\gg $T$_2$ as in the Levstein proposal.  Numerical simulations along with experiments on C$_{60}$, C$_{70}$, and adamantane were performed to verify these theoretical predictions.

  The paper is organized as follows. {\bf{Section II}} presents the theoretical description of long-lived echoes under an imperfect CPMG($\phi_{1},\,\phi_{2}$) pulse train along with numerical spin simulations in a linear, 10 spin chain.  In {\bf{Section III}}, a brief description of the experimental setup and the chemical systems that were studied is provided.  In {\bf{Section IV}}, the experimental results are presented and interpreted with respect to the theoretical framework and numerical simulations presented in {\bf{Section II}}.  In {\bf{Section V}}, conclusions and a summary of this work are presented, along with an explicit comparison of our proposed theory to both the Barrett\cite{Li07,Li08,Dong08} and Levstein\cite{Franzoni05,Franzoni08,Levstein08,Franzoni12} proposals described above.  In {\bf{Appendix A}}, explicit details of average Hamiltonian calculations performed in {\bf{Section II}} are provided, and in {\bf{Appendix B}},  the effects of both $\widehat{H}_{cs}$ and $\widehat{H}_{D}$ during the $\pi-$pulses to the long-lived echoes are presented.

\section{Theory and Numerical Simulations}
Consider a system of $N$ homonuclear, $I=1/2$ spins interacting via the dipolar interaction and placed in a strong magnetic field aligned about the $\widehat{z}$-axis.  In the high-field limit, the dipolar Hamiltonian can be written in the rotating frame of the Zeeman interaction as:
\begin{eqnarray}
\frac{\widehat{H}_{D}}{\hbar}&=&\sum_{j<k}\omega_{D}^{jk}\left(2\widehat{I}_{Z,j}\widehat{I}_{Z,k}-\widehat{I}_{X,j}\widehat{I}_{X,k}-\widehat{I}_{Y,j}\widehat{I}_{Y,k}\right)
\label{eq:Hdip}
\end{eqnarray}
where $\omega_{D}^{jk}=\frac{\gamma^{2}}{r_{jk}^{3}}\frac{3\cos^{2}(\theta_{jk})-1}{2}$ is the dipolar coupling constant between spins $j$ and $k$, $r_{jk}=|\vec{r}_{jk}|=\left|\vec{r}_{j}-\vec{r}_{k}\right|$ is the magnitude of the internuclear vector between spins $j$ and $k$, $\theta_{jk}=\arccos\left(\frac{\vec{r}_{jk}\cdot\widehat{z}}{r_{jk}}\right)$ is the angle $\vec{r}_{jk}$ makes with the applied magnetic field.  $\widehat{I}_{Z,k}$, $\widehat{I}_{X,k}$, and $\widehat{I}_{Y,k}$ are the spin-1/2 operators for spin $k$. Since $[\widehat{H}_{D},\widehat{I}_{Z}]=0$, where $\widehat{I}_{Z}=\sum_{j=1}^{N}\widehat{I}_{Z,j}$ is the total spin magnetization along the $\widehat{z}$-direction, simultaneous orthonormal eigenstates of both $\widehat{H}_{D}$ and $\widehat{I}_{Z}$ can be found, which are denoted by $|\epsilon_{k,n}\rangle$ where $\widehat{H}_{D}|\epsilon_{k,n}\rangle=\epsilon_{k,n}|\epsilon_{k,n}\rangle$, $\widehat{I}_{Z}|\epsilon_{k,n}\rangle=n|\epsilon_{k,n}\rangle$, and $\langle \epsilon_{k,n}|\epsilon_{j,m}\rangle=\delta_{kj}\delta_{mn}$.  In this case, both $\widehat{H}_{D}$ and $\widehat{I}_{Z}$ can be written in this eigenbasis as
 $\widehat{H}_{D}=\sum_{(k,n)}\epsilon_{k,n}|\epsilon_{k,n}\rangle\langle \epsilon_{k,n}|$ and $\widehat{I}_{Z}=\sum_{(k,n)}n|\epsilon_{k,n}\rangle\langle\epsilon_{k,n}|$.  One important symmetry of $\widehat{H}_{D}$ is its invariance to $\pi-$rotations about the transverse plane, i.e., $\widehat{P}_{\frac{\pi}{2},\phi}(\pi)\widehat{H}_{D}\widehat{P}^{\dagger}_{\frac{\pi}{2},\phi}(\pi)=\widehat{H}_{D}$ where  \begin{eqnarray}
 \widehat{P}_{\theta,\phi}(\Theta)=e^{-i\Theta\left(\cos(\theta)\widehat{I}_{Z}+\sin(\theta)\left(\widehat{I}_{X}\cos(\phi)+\widehat{I}_{Y}\sin(\phi)\right)\right)}
 \label{eq:pulse}
  \end{eqnarray}
  represents a ``pure'' rotation of $\Theta$ about an axis defined by spherical coordinates $(\theta,\phi)$ with $\theta\in [0,\pi]$ and $\phi\in[0,2\pi)$.

Aside from the dipolar interaction, another important interaction for $I=1/2$ nuclei is the chemical shift, $\widehat{H}_{cs}$, given by:
\begin{eqnarray}
\frac{\widehat{H}_{cs}}{\hbar}&=&\sum_{j=1}^{N}\omega_{cs}^{j}\widehat{I}_{Z,j}=\sum_{k,m}\left(H_{cs}\right)^{m,m}_{k,k}|\epsilon_{k,m}\rangle\langle\epsilon_{k,m}|\nonumber\\
&+&\sum_{(k<j),m}
\left(H_{cs}\right)^{m,m}_{j,k}|\epsilon_{j,m}\rangle\langle\epsilon_{k,m}|+\left(H_{cs}\right)^{m,m}_{k,j}|\epsilon_{k,m}\rangle\langle\epsilon_{j,m}|
\end{eqnarray}
where $\left(H_{cs}\right)^{a,b}_{j,k}=\left\langle \epsilon_{j,a}\left|\frac{\widehat{H}_{cs}}{\hbar}\right|\epsilon_{k,b}\right\rangle$.  In general, $\omega_{cs}^{j}$ is anisotropic and depends upon crystallite orientation.  Unlike $\widehat{H}_{D}$, $\widehat{H}_{cs}$ is antisymmetric with respect to $\pi-$rotations about the transverse plane, i.e., $\widehat{P}_{\frac{\pi}{2},\phi}(\pi)\widehat{H}_{cs}\widehat{P}_{\frac{\pi}{2},\phi}^{\dagger}(\pi)=-\widehat{H}_{cs}$.

 From Fig. \ref{fig:fig1}(B), the propagator for one loop of a CPMG($\phi_{1},\phi_{2}$) pulse train ($n_{l}=1$) with perfect $\pi-$pulses is given by:
 \begin{eqnarray}
 \widehat{U}^{ideal}_{\text{CPMG}_{\phi_{1},\phi_{2}}}(4\tau)=\widehat{U}_{f}(\tau)\widehat{P}_{\frac{\pi}{2},\phi_{2}}(\pi)\widehat{U}_{f}(2\tau)\widehat{P}_{\frac{\pi}{2},\phi_{1}}(\pi)\widehat{U}_{f}(\tau)
 \label{eq:cpmgideal}
 \end{eqnarray}
  where $\widehat{U}_{f}(t)=e^{-i\frac{t}{\hbar}\left(\widehat{H}_{D}+\hbar\omega_{\text{off}}\widehat{I}_{Z}+\widehat{H}_{cs}\right)}$ with $\omega_{\text{off}}$ being some global resonance offset.  With perfect $\pi-$pulses in Eq. (\ref{eq:cpmgideal}), no long-lived echoes are predicted (as shown from simulations), and if $\widehat{H}_{cs}=0$, the effects of a CPMG($\phi_{1},\phi_{2}$) pulse train is identical to that from the corresponding Hahn echo experiments in dipolar solids.  As mentioned in the introduction, previous theories for observing long-lived echoes require  imperfect $\pi-$pulses, either through consideration of dipolar evolution during the pulse\cite{Li07,Li08} or due to field inhomogeneites that render the applied pulse imperfect\cite{Franzoni05,Franzoni08,Levstein08}.
While including $\widehat{H}_{D}$ during the pulse would definitely be needed if the interpulse spacing, $2\tau$, was comparable to the $\pi-$pulse time, $t_{p}=\frac{\pi}{\omega_{RF}}$ where $\omega_{RF}$ is the strength of the applied RF field, in cases where $2\tau\gg t_{p}$ (which is the regime that was experimentally studied in this work) and where $\widehat{H}_{D}t_{p}\ll 1$, neglecting $\widehat{H}_{D}$ during the pulse appears to be justified [the effects of $\widehat{H}_{D}$ during the $\pi-$pulse were shown in {\bf{Appendix B}} to not significantly alter the predictions of the following theory when $\tau\gg t_{p}$].  Therefore, in this work we consider the case where the $\pi-$pulses are ``imperfect'' as a result of resonance offsets during the applied pulse [$\theta\neq \frac{\pi}{2}$ in Eq. (\ref{eq:pulse})] and/or small flip-angle errors due to $\pi-$pulse miscalibration [$(\omega_{RF}+\delta\omega_{RF})t_{p}=\Theta\neq \pi$ in Eq. (\ref{eq:pulse})].  An ``imperfect'' $\pi-$pulse, $\widehat{R}_{\phi}(\pi)$, can be written as:
\begin{eqnarray}
\widehat{R}_{\phi}(\pi)&=&e^{-it_{p}\left[(\omega_{RF}+\delta\omega_{RF})\left(\widehat{I}_{X}\cos(\phi)+\widehat{I}_{Y}\sin(\phi)\right)+\omega_{\text{off}}\widehat{I}_{Z}\right]}
\label{eq:impii}
\end{eqnarray}
where $\delta\omega_{RF}$ and $\omega_{\text{off}}$ are due to imperfect $\pi-$pulse calibration and resonance offset, respectively.  If $\left|\frac{\omega_{
\text{off}}}{\omega_{RF}}\right|\ll 1$ and $\left|\frac{\delta\omega_{RF}}{\omega_{RF}}\right|\ll 1$, AHT\cite{Haeberlen68} can be used to approximate $\widehat{R}_{\phi}(\pi)$ in Eq. (\ref{eq:impii}) as
\begin{eqnarray}
\widehat{R}_{\phi}(\pi)&\approx&\widehat{P}_{\frac{\pi}{2},\phi}(\pi)e^{-it_{p}\left[\delta\omega_{RF}\left(\widehat{I}_{X}\cos(\phi)+\widehat{I}_{Y}\sin(\phi)\right)+\frac{2\omega_{\text{off}}}{\pi}\left(\widehat{I}_{Y}\cos(\phi)-\widehat{I}_{X}\sin(\phi)\right)\right]}
\nonumber\\
&=&\widehat{P}_{\frac{\pi}{2},\phi}(\pi)e^{-i\frac{t_{p}}{2}\left[\left(\delta\omega_{RF}-\frac{2i\omega_{\text{off}}}{\pi}\right)\widehat{I}_{+}e^{-i\phi}+\left(\delta\omega_{RF}+\frac{2i\omega_{\text{off}}}{\pi}\right)\widehat{I}_{-}e^{i\phi}\right]}\nonumber\\
&=&\widehat{P}_{\frac{\pi}{2},\phi}(\pi)e^{-i\frac{t_{p}}{2}\left[\sum_{(j,k,m)}\delta\omega e^{-i(\phi+\delta\phi)}\left(\widehat{I}_{+}\right)^{m,m-1}_{k,j}|\epsilon_{k,m}\rangle\langle\epsilon_{j,m-1}|+\delta\omega e^{i(\phi+\delta\phi)}\left(\widehat{I}_{-}\right)^{m-1,m}_{j,k}|\epsilon_{j,m-1}\rangle\langle \epsilon_{k,m}|\right]}\nonumber\\
\label{eq:impi}
\end{eqnarray}
where $\delta\omega=\sqrt{\delta\omega_{RF}^{2}+\frac{4}{\pi^{2}}\omega_{\text{off}}^{2}}$, $e^{-i\delta\phi}=\frac{\delta\omega_{RF}-\frac{2i}{\pi}\omega_{\text{off}}}{\delta\omega}$, $\widehat{I}_{\pm}=\widehat{I}_{X}\pm i\widehat{I}_{Y}$, and $\left(\widehat{I}_{\pm}\right)^{a,b}_{j,k}=\langle \epsilon_{j,a}|\widehat{I}_{\pm}|\epsilon_{k,b}\rangle$.  Details of the calculations in Eq. (\ref{eq:impi}) are provided in {\bf{Appendix A}}.  Also, the case where evolution under $\widehat{H}_{cs}$ during the $\pi-$pulse in Eq. (\ref{eq:impi}) is also provided in {\bf{Appendix B}}.  Extensions to the case where $\delta\omega_{RF}$ and $\omega_{\text{off}}$ are spatially-dependent can be readily performed.    Note that if the pulse error is completely due to resonance offsets ($\delta\omega_{RF}=0$), $\delta\phi=\pm \frac{\pi}{2}$, whereas if it is entirely due to flip-angle error ($\omega_{\text{off}}=0$), $\delta\phi=0$ or $\delta\phi=\pi$.

We now consider the analysis of a single CPMG$(\phi_{1},\phi_{2})$ pulse train ($n_{l}=1$) using imperfect $\pi-$pulses, $\widehat{R}_{\phi}(\pi)$ in Eq. (\ref{eq:impi}).  In this case, the propagator is given by $\widehat{U}_{\text{CPMG}_{\phi_{1},\phi_{2}}}(4\tau)=\widehat{U}_{f}(\tau)\widehat{R}_{\phi_{2}}(\pi)\widehat{U}_{f}(2\tau)\widehat{R}_{\phi_{1}}(\pi)\widehat{U}_{f}(\tau)$.  Transforming into an interaction frame of $\widehat{H}_{D}+\hbar\omega_{\text{off}}\widehat{I}_{Z}$ defined by $\widehat{U}(t)=e^{-\frac{it}{\hbar}(\widehat{H}_{D}+\hbar\omega_{\text{off}}\widehat{I}_{Z})}$,  $\widehat{U}_{\text{CPMG}_{\phi_{1},\phi_{2}}}(4\tau)$ can be approximated as (with details of the calculations provided in Appendix A):
\begin{eqnarray}
\widehat{U}_{\text{CPMG}_{\phi_{1},\phi_{2}}}(4\tau)\approx \widehat{P}_{0,0}(2(\phi_{2}-\phi_{1}))\widehat{U}_{D}(2\tau)e^{-i\frac{2t_{p}}{\hbar}\overline{H}_{\text{avg}}}\widehat{U}_{D}(2\tau)
\label{eq:cpmg1aloop}
\end{eqnarray}
where $\widehat{U}_{D}(t)=e^{-\frac{it}{\hbar}\widehat{H}_{D}}$ is the propagator under pure dipolar evolution, and $\overline{H}_{\text{avg}}$ is the average Hamiltonian in the interaction frame of $\widehat{H}_{D}+\hbar\omega_{\text{off}}\widehat{I}_{Z}$ for the CPMG($\phi_{1},\phi_{2}$) pulse train with $n_l=1$.  The average Hamiltonian can be written as $\overline{H}_{\text{avg}}=\sum_{k=1}^{\infty}\overline{H}_{\text{avg}}^{(k)}$ where $\overline{H}_{\text{avg}}^{(k)}$ is the $k^{th}$-order contribution to $\overline{H}_{\text{avg}}$.  The first-order contribution, $\overline{H}_{\text{avg}}^{(1)}$, written in the simultaneous eigenbasis of $\widehat{H}_{D}$ and $\widehat{I}_{Z}$ is given by:
{\small{\begin{eqnarray}
\frac{\overline{H}_{\text{avg}}^{(1)}}{\hbar}&=&\sum_{(k,m),(j,m-1)}\overline{\lambda}^{m,m-1}_{kj}|\epsilon_{k,m}\rangle\langle\epsilon_{j,m-1}|+\left(\overline{\lambda}^{m,m-1}_{kj}\right)^{*}|\epsilon_{j,m-1}\rangle\langle\epsilon_{k,m}|\nonumber\\
&+&\frac{2\tau}{t_{p}}\sum_{(k,m)<(l,m)}\left(\text{sinc}[2\omega^{m,m}_{kl}\tau]-\text{sinc}[\omega^{m,m}_{kl}\tau]\right)\left(\left(\widehat{H}_{cs}\right)^{m,m}_{k,l}|\epsilon_{k,m}\rangle\langle\epsilon_{l,m}|
+\left(\widehat{H}_{cs}\right)^{m,m}_{l,k}|\epsilon_{l,m}\rangle\langle\epsilon_{k,m}|\right)\nonumber\\
\label{eq:Havg1}
\end{eqnarray}}}
where $\omega^{m,n}_{kj}=\frac{\epsilon_{k,m}-\epsilon_{j,n}}{\hbar}$,
{\footnotesize{
\begin{eqnarray}
\overline{\lambda}^{m,n}_{kj}&=&\frac{\delta\omega}{2}\cos\left(\omega^{m,n}_{kj}\tau+\Delta\chi\right)e^{i\Psi_{1}}\left(\widehat{I}_{+}\right)^{m,n}_{k,j}\nonumber\\
\label{eq:amplo}
\end{eqnarray}}}
 where $\Delta\chi=\frac{\phi_{2}-\phi_{1}}{2}+\delta\phi-\omega_{\text{off}}\tau$ and $\Psi_{1}=\frac{\phi_{2}-3\phi_{1}}{2}$.  The contributions of $\widehat{H}_{cs}$ to $\overline{H}^{(1)}_{\text{avg}}$ in Eq. (\ref{eq:Havg1}), which are due to the fact that typically $\left[\widehat{H}_{D},\,\widehat{H}_{cs}\right]\neq 0$, contain only zero-quantum transitions between eigenstates of $\widehat{H}_{D}$ and are therefore not the source of the long-lived single-quantum spin echoes observed under CPMG.  In the following, we therefore neglect the effects of $\widehat{H}_{cs}$ during both $\tau$ and $t_{p}$, although explicit calculations of the effects of $\widehat{H}_{cs}$ on the long-lived spin echoes are presented in {\bf{Appendix B}}.   Neglecting these contributions of $\widehat{H}_{cs}$, Eq. (\ref{eq:Havg1}) can be written in terms of spin operators as:
 \begin{eqnarray}
\frac{ \overline{H}^{(1)}_{\text{avg}}}{\hbar}&=&\frac{\delta\omega}{4}\left(\widehat{U}_{D}(\tau)\widehat{I}_{+}\widehat{U}^{\dagger}_{D}(\tau)e^{-i\Delta\chi}+\widehat{U}^{\dagger}_{D}(\tau)\widehat{I}_{+}\widehat{U}_{D}(\tau)e^{i\Delta\chi}\right)e^{i\Psi_{1}}\nonumber\\
&+&\frac{\delta\omega}{4}\left(\widehat{U}_{D}(\tau)\widehat{I}_{-}\widehat{U}^{\dagger}_{D}(\tau)e^{i\Delta\chi}+\widehat{U}^{\dagger}_{D}(\tau)\widehat{I}_{-}\widehat{U}_{D}(\tau)e^{-i\Delta\chi}\right)e^{-i\Psi_{1}}\nonumber\\
&=&\frac{\delta\omega}{2}\widehat{U}_{D}(\tau)\left(\widehat{I}_{X}\cos\left(\Psi_{1}-\Delta\chi\right)-\widehat{I}_{Y}\sin\left(\Psi_{1}-\Delta\chi\right)\right)\widehat{U}^{\dagger}_{D}(\tau)\nonumber\\
&+&\frac{\delta\omega}{2}\widehat{U}^{\dagger}_{D}(\tau)\left(\widehat{I}_{X}\cos\left(\Psi_{1}+\Delta\chi\right)-\widehat{I}_{Y}\sin\left(\Psi_{1}+\Delta\chi\right)\right)\widehat{U}_{D}(\tau)\nonumber\\
\label{eq:avgH1}
\end{eqnarray}

The contributions of pulse-flip errors and offsets to $\overline{H}_{\text{avg}}^{(2)}$, written in terms of spin operators and once again neglecting the effects of $\widehat{H}_{cs}$, is given by:
{\footnotesize{
\begin{eqnarray}
\frac{\overline{H}_{\text{avg}}^{(2)}}{\hbar}&=&\frac{\delta\omega^{2}t_{p}}{16i}\left(\widehat{U}^{\dagger}_{D}(\tau)\widehat{I}_{+}\left(\widehat{U}_{D}(\tau)\right)^{2}\widehat{I}_{+}\widehat{U}_{D}^{\dagger}(\tau)-\widehat{U}_{D}(\tau)\widehat{I}_{+}\left(\widehat{U}^{\dagger}_{D}(\tau)\right)^{2}\widehat{I}_{+}\widehat{U}_{D}(\tau)\right)e^{2i\Psi_{1}}\nonumber\\
&+&\frac{\delta\omega^{2}t_{p}}{16i}\left(\widehat{U}^{\dagger}_{D}(\tau)\widehat{I}_{-}\left(\widehat{U}_{D}(\tau)\right)^{2}\widehat{I}_{-}\widehat{U}_{D}^{\dagger}(\tau)-\widehat{U}_{D}(\tau)\widehat{I}_{-}\left(\widehat{U}^{\dagger}_{D}(\tau)\right)^{2}\widehat{I}_{-}\widehat{U}_{D}(\tau)\right)e^{-2i\Psi_{1}}\nonumber\\
&+&\frac{\delta\omega^{2}t_{p}}{16i}\left(\widehat{U}^{\dagger}_{D}(\tau)\widehat{I}_{+}\left(\widehat{U}_{D}(\tau)\right)^{2}\widehat{I}_{-}\widehat{U}_{D}^{\dagger}(\tau)-\widehat{U}_{D}(\tau)\widehat{I}_{-}\left(\widehat{U}^{\dagger}_{D}(\tau)\right)^{2}\widehat{I}_{+}\widehat{U}_{D}(\tau)\right)e^{i2\Delta\chi}\nonumber\\
&+&\frac{\delta\omega^{2}t_{p}}{16i}\left(\widehat{U}^{\dagger}_{D}(\tau)\widehat{I}_{-}\left(\widehat{U}_{D}(\tau)\right)^{2}\widehat{I}_{+}\widehat{U}_{D}^{\dagger}(\tau)-\widehat{U}_{D}(\tau)\widehat{I}_{+}\left(\widehat{U}^{\dagger}_{D}(\tau)\right)^{2}\widehat{I}_{-}\widehat{U}_{D}(\tau)\right)e^{-i2\Delta\chi}\nonumber\\
\label{eq:Havg2}
\end{eqnarray}}}
In this case, a second-order energy shift for each state $|\epsilon_{k,m}\rangle$ occurs along with additional zero-quantum transitions between states $|\epsilon_{k,m}\rangle$ and $|\epsilon_{l,m}\rangle$ and double-quantum transitions between the states $|\epsilon_{k,m}\rangle$ and $|\epsilon_{j,m\pm 2}\rangle$.  We now rewrite the propagator for the CPMG($\phi_{1},\phi_{2}$) pulse block ($n_{l}=1$) in Eq. (\ref{eq:cpmg1aloop}) in a more suggestive form as:
\begin{eqnarray}
\widehat{U}_{\text{CPMG}_{\phi_{1},\phi_{2}}}(4\tau)\approx \widehat{P}_{0,0}(2(\phi_{2}-\phi_{1}))\widehat{U}_{D}(2\tau)\widetilde{R}^{\text{eff.}}_{\Psi_{1}}\widehat{U}_{D}(2\tau)
\label{eq:cpmg1loop}
\end{eqnarray}
where $\widetilde{R}^{\text{eff.}}_{\Psi_{1}}\equiv e^{-i\frac{2t_{p}}{\hbar}\overline{H}_{\text{avg}}}$
denotes an effective excitation of phase $\Psi_{1}$.  From Eq. (\ref{eq:cpmg1loop}), the propagator under one loop ($n_{l}=1$) is equivalent to a mostly single-quantum excitation (if $||\overline{H}_{\text{avg}}^{(1)}||\gg ||\overline{H}_{\text{avg}}^{(2)}||$ where $||A||=\sqrt{\text{Trace}\left[A^{\dagger}A\right]}$ represents the Frobenius matrix norm) of phase $\Psi_{1}$, $\widetilde{R}^{\text{eff.}}_{\Psi_{1}}$, applied between
periods of free evolution under $\widehat{H}_{D}$ for times $2\tau$.

   If $\omega_{kj}^{m,m-1}\tau\ll 1$ for all $(k,m)$ and $(j,m-1)$, $\widehat{R}^{\text{eff.}}_{\Psi_{1}}\approx \widehat{P}_{\frac{\pi}{2},-\Psi_{1}}(2\delta\omega t_{p}\cos(\Delta\chi))$, which can also be seen from $\overline{H}^{(1)}_{\text{avg}}$ in Eq. (\ref{eq:avgH1}) and $\overline{H}_{\text{avg}}^{(2)}$ in Eq. (\ref{eq:Havg2}) by approximating $\widehat{U}_{D}(\tau)=e^{-i\frac{\tau}{\hbar}\widehat{H}_{D}}\approx \widehat{1}$:
\begin{eqnarray}
\frac{\overline{H}^{(1)}_{\text{avg}}}{\hbar}&=&\delta\omega\cos(\Delta\chi)\left(\widehat{I}_{X}\cos\left(\Psi_{1}\right)-\widehat{I}_{Y}\sin\left(\Psi_{1}\right)\right)\nonumber\\
\frac{\overline{H}^{(2)}_{\text{avg}}}{\hbar}&=&\frac{\delta\omega^{2}t_{p}}{4}\sin(2\Delta\chi)\widehat{I}_{Z}
\label{eq:Happrox}
\end{eqnarray}
In this case, $\widetilde{R}^{\text{eff.}}_{\Psi_{1}}$ is equivalent to an off resonant RF pulse of strength $\omega^{\text{eff.}}_{RF}=\delta\omega \cos(\Delta\chi)$ and phase $-\Psi_{1}$, applied off-resonantly with $\omega^{\text{eff.}}_{\text{off.}}=\frac{\delta\omega^{2}t_{p}}{4}\sin(2\Delta\chi)$ for a time $2t_{p}$.  However, when
 $\omega_{jk}^{m,m-1}\tau \geq 1$, $\overline{H}_{\text{avg}}$ no longer behaves like an RF pulse and is attenuated by the transition frequencies,  $\omega_{kj}^{m,m-1}$, and by additional Bloch-Siegert\cite{Bloch40} offset terms that arise from $\overline{H}^{(2)}_{\text{avg}}$ in Eq. (\ref{eq:Havg2}).

For $n_{l}$ repetitions of the $\text{CPMG}(\phi_{1},\phi_{2})$ pulse block, the propagator can be rewritten as:
\begin{eqnarray}
\widehat{U}_{\text{CPMG}_{\phi_{1},\phi_{2}}}(4n_{l}\tau)&=&\widehat{P}_{0,0}(2n_{l}(\phi_{2}-\phi_{1}))T\prod_{k=1}^{n_{l}}\left(\widehat{U}_{D}(2\tau)\widetilde{R}^{\text{eff.}}_{\Psi_{k}}\widehat{U}_{D}(2\tau)\right)\nonumber\\
\label{eq:cpmgnloop}
\end{eqnarray}
where $T$ is the Dyson time ordering operator, and $\Psi_{k}=\Psi_{1}+2(k-1)(\phi_{2}-\phi_{1})$ is the effective phase of the $k^{th}$ excitation pulse.  In this work, we will restrict our analysis to the CPMG($X,\,X$) $\equiv$ CPMG$(0,\,0)$ [$\Psi_{1}=0$], CPMG($Y,\,Y$) $\equiv$ CPMG$\left(\frac{\pi}{2},\,\frac{\pi}{2}\right)$ $\left[\Psi_{1}=-\frac{\pi}{2}\right]$, CPMG($X,-X$) $\equiv$ CPMG$(0,\,\pi)$ $\left[\Psi_{1}=\frac{\pi}{2}\right]$ and CPMG ($Y,-Y$) $\equiv$ CPMG$\left(\frac{\pi}{2},\,\frac{3\pi}{2}\right)$ $\left[\Psi_{1}=0\right]$ sequences, in which case, $\widehat{P}_{0,0}(2(\phi_2-\phi_1))=\widehat{1}$ in Eq. (\ref{eq:cpmgnloop}) and $\text{mod}\left[\Psi_{k},\,2\pi\right]=\Psi_{1}$ for all $k$.  For these CPMG sequences, the following propagators are obtained:
\begin{eqnarray}
\widehat{U}_{\text{CPMG}_{Y,\,Y}}&=&\prod_{k=1}^{n_{l}}\left(\widehat{U}_{D}(2\tau)\widetilde{R}^{\text{eff.}}_{-Y}\widehat{U}_{D}(2\tau)\right)\nonumber\\
\widehat{U}_{\text{CPMG}_{X,-X}}&=&\prod_{k=1}^{n_{l}}\left(\widehat{U}_{D}(2\tau)\widetilde{R}^{\text{eff.}}_{Y}\widehat{U}_{D}(2\tau)\right)\nonumber\\
\widehat{U}_{\text{CPMG}_{Y,-Y}}&=&\prod_{k=1}^{n_{l}}\left(\widehat{U}_{D}(2\tau)\widetilde{R}^{\text{eff.}}_{X}\widehat{U}_{D}(2\tau)\right)\nonumber\\
\widehat{U}_{\text{CPMG}_{X,X}}&=&\prod_{k=1}^{n_{l}}\left(\widehat{U}_{D}(2\tau)\widetilde{R}^{\text{eff.}}_{X}\widehat{U}_{D}(2\tau)\right)\nonumber\\
\label{eq:cpmgprops}
\end{eqnarray}
Equation (\ref{eq:cpmgprops}) represents the main theoretical result of this work.  The propagators in Eq. (\ref{eq:cpmgprops}) resemble a series of single-quantum excitations of either phase $\pm Y$ for the CPMG($Y,\,Y$) and CPMG($X,-X$) pulse trains, respectively, or of phase $X$ for the CPMG($X,\,X$) and CPMG($Y,-Y$) pulse trains, that are separated by a time $4\tau$ during which evolution under $\widehat{H}_{D}$ occurs.

 Before application of the CPMG pulse train, a $\left(\frac{\pi}{2}\right)_{X}$ pulse is applied [Fig. \ref{fig:fig1}(B)] that rotates the initial equilibrium $\widehat{z}$-magnetization, $\widehat{\rho}_{eq}=\widehat{I}_{Z}$, into transverse magnetization:
 \begin{eqnarray}
 \widehat{\rho}(0)&=&-\widehat{I}_{Y}=\frac{i}{2}\left(\widehat{I}_{+}-\widehat{I}_{-}\right)=\frac{i}{2}\sum_{(k,m)<(j,m-1)}\left(I_{+}\right)^{m,m-1}_{k,j}|\epsilon_{k,m}\rangle\langle\epsilon_{j,m-1}|-\left(I_{-}\right)_{j,k}^{m-1,m}|\epsilon_{j,m-1}\rangle\langle\epsilon_{k,m}|\nonumber\\
 \label{eq:ymag}
 \end{eqnarray}
 The phase of the coherence between the states $|\epsilon_{k,m}\rangle$ and $|\epsilon_{j,m-1}\rangle$ in Eq. (\ref{eq:ymag}) is $\phi=\frac{3\pi}{2}$, i.e., $-Y$.  From Eq. (\ref{eq:cpmgprops}), both the CPMG($Y,\,Y$) and CPMG($X,-X$) pulse trains resemble the propagators found under pulsed spin-locking sequences of initial $\pm \widehat{y}$-magnetization in dipolar solids\cite{Suwelack80,Ivanov78,Maricq85}. It is thus predicted that the CPMG($Y,\,Y$) and CPMG($X,-X$) sequences can generate an effective pulsed spin-locking of initial $\widehat{y}$-magnetization, whereas the CPMG($X,\,X$) and CPMG($Y,-Y$) sequences cannot.

While the above arguments suggest that the CPMG($Y,\,Y$) and CPMG($X,-X$) sequences will spin-lock those coherences with phase $\pm$Y, the strength of the spin-locking field/effective pulse is different for the CPMG($Y,\,Y$) and CPMG($X,-X$) sequences and depends upon the type of ``error'' introduced into the imperfect $\pi-$pulses, resonance offsets vs. pulse flip error.  From Eq. (\ref{eq:amplo}), the strength of the effective field under the CPMG($Y,\,Y$) pulse train for transitions that satisfy $4\omega_{jk}^{m,m-1}\tau=2\pi n$ for integer $n$ is proportional to $\cos\left(\frac{n\pi}{2}+\delta\phi-\omega_{\text{off}}\tau\right)$ whereas it is proportional to $\sin\left(\frac{n\pi}{2}+\delta\phi-\omega_{\text{off}}\tau\right)$ for the CPMG($X,-X$) sequence.  When there is only pulse flip error and $\omega_{\text{off}}=0$, $\delta\phi=0$ or $\delta\phi=\pi$.   In this case, the effective field is nonzero for the CPMG($Y,\,Y$) pulse train when $n$ is even and is nonzero for the  CPMG($X,-X$) pulse train when $n$ is odd.  If $\omega_{\text{off}}\neq 0$, then the effective field is nonzero for both the CPMG($Y,\,Y$) and CPMG($X,-X$) pulse trains, regardless of $n$ .  When there is only a resonance offset, i.e., the pulse is applied off-resonance with $\omega_{\text{off}}\neq 0$ but with $\delta\omega_{RF}=0$, $\delta\phi=\pm\frac{\pi}{2}$, and the spin-locking field/effective pulse for both sequences will in general be nonzero with the assumption that $\omega_{\text{off}}\tau\neq m\pi$ for integer $m$.

To illustrate the above predictions, numerical simulations of the CPMG sequences were performed on a ``toy model'' system consisting of 10 spin-1/2 particles linearly arranged along the $\widehat{z}$ axis, where the position of the $k^{th}$ spin was initially given by $\vec{r}_{k}=k\widehat{z}$.  The dipolar coupling constant between spins $k$ and $j$ was given by $\frac{\omega_{D}^{jk}}{2\pi}=\frac{10\,\text{Hz}}{|\vec{r}_{j}-\vec{r}_{k}|^{3}}$.  In order to mimic the effects of random spin distributions in solids (due to the $1\%$ natural abundance of $^{13}$C spins) and to generate a ``quasi''-continuous spectrum, the $\widehat{z}$-coordinate of the $k^{th}$ spin was randomly chosen to be within the interval $[k-0.1, k+0.1]$ for all $k$.  The dynamics under the various CPMG pulse trains were then averaged over 1250 different random configurations of the spin system, which resulted in a ``quasi''-continuous spectrum with a dipolar linewidth of approximately 40 Hz.  In all simulations, $\frac{\omega_{RF}}{2\pi}=2$ kHz, and the pulses were simulated as ``real" pulses, i.e., dipolar evolution was considered during the pulse, although neglecting $\widehat{H}_{D}$ during the pulse did not significantly affect the results of the simulations under the conditions that were studied.  Relaxation effects were not included in the simulations.  Finally, it should be noted that an even number of spins was chosen to avoid single-quantum constants of motion\cite{Walls06} that are due to the invariance of $\widehat{H}_{D}$ under $\pi$ rotations about the transverse plane.

\begin{figure}
\includegraphics*[scale=.5]{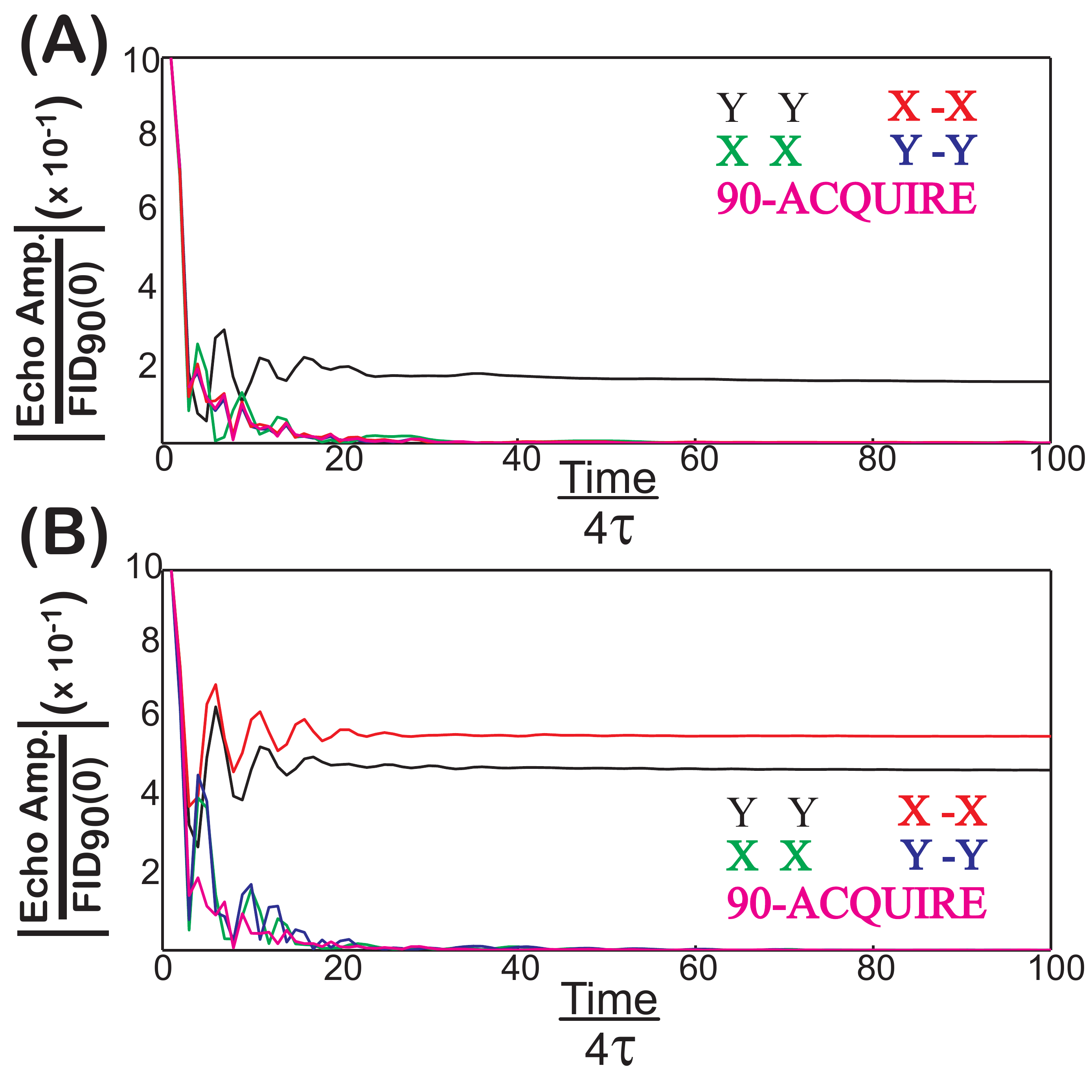}
\caption{(Color online) Numerical simulations of the (black curve) CPMG($Y,\,Y$), (green curve) CPMG($X,\,X$), (red curve) CPMG($X,-X$), and (blue curve) CPMG($Y,-Y$) pulse trains applied to a system of 10 spins, averaged over 1250 different linear arrangements of the spins.  The absolute value of $\left|\overline{\text{Trace}\left[ I_{+}\widehat{U}_{\text{CPMG}}(4n\tau)\widehat{I}_{Y}\widehat{U}^{\dagger}_{\text{CPMG}}(4n\tau)\right]}\right|/\text{Trace}\left[\left(\widehat{I}_{Y}\right)^{2}\right]$ at integer multiples of 4$\tau$ ($\tau=2.9$ ms), where $\widehat{U}_{\text{CPMG}}(4n\tau)$ is the exact propagator for the CPMG($\phi_{1},\phi_{2}$) sequence.  The FID evaluated at integer multiples of 4$\tau$, $\left|\overline{\text{Trace}\left[\widehat{I}_{+}\widehat{U}_{D}(4n\tau)\widehat{I}_{Y}\widehat{U}^{\dagger}_{D}(4n\tau)\right]}\right|/\text{Trace}\left[\left(\widehat{I}_{Y}\right)^{2}\right]$, is also shown for comparison (magenta curve).  In (A), imperfect, on resonant $\pi-$pulses were used in the CPMG sequences with a flip angle of $(\omega_{RF}+\delta\omega_{RF})t_{p}=\frac{37}{36}\pi$ [$185^{\circ}$ pulses vs. $180^{\circ}$ pulses].  As predicted from Eq. (\ref{eq:amplo}), only the CPMG($Y,\,Y$) sequence leads to any long-lived echoes (black curve) under flip-angle errors, and a clear quasiequilibrium containing nonzero $\widehat{y}$-magnetization was observed.  In (B), pulses with $\omega_{RF}t_{p}=\pi$ ($\delta\omega_{RF}=0$) were applied but with a resonance offset of $\frac{\omega_{\text{off}}}{2\pi}=300$Hz.  In this case, both the (black curve) CPMG($Y,\,Y$) and (red curve) CPMG($X,-X$) generated long-lived echoes.  In both (A) and (B), the echo amplitudes under both the (green curve) CPMG($X,\,X$) and (blue curve) CPMG($Y,-Y$) pulse trains were found to decay on a time comparable to the FID from a $\left(\frac{\pi}{2}\right)_{X}-$acquire (magenta curve).
}
\label{fig:fig2}
\end{figure}

Figure \ref{fig:fig2} shows a simulation of the magnitude of the echo amplitudes, $\left|\overline{\text{Trace}\left[\widehat{I}_{+}\widehat{U}_{\text{CPMG}}(t)\widehat{I}_{Y}\widehat{U}^{\dagger}_{\text{CPMG}}(t)\right]}\right|$ where $\widehat{U}_{\text{CPMG}}(t)$ is the exact propagator under the CPMG($\phi_{1},\phi_{2}$) sequence, evaluated at integer multiples of $4\tau$ ($\tau=2.9$ ms) and averaged over 1250 different linear arrangements of a 10 spin system, evolving under the (black curves) CPMG($Y,\,Y$), (red curves) CPMG($X,-X$), (green curves) CPMG($X,\,X$) and (blue curves) CPMG($Y,-Y$) pulse trains.  The corresponding FID from a $\left(\frac{\pi}{2}\right)_{X}-$acquire simulation evaluated at integer multiples of $4\tau$, \\
$\left[\text{magenta curves},\,\left|\overline{\text{Trace}\left[\widehat{I}_{+}\widehat{U}_{D}(4n\tau)\widehat{I}_{Y}\widehat{U}^{\dagger}_{D}(4n\tau)\right]}\right|\right]$, is shown for comparison.  In Fig. \ref{fig:fig2}(A), the effect of a pulse flip error of $5^{\circ}$ degrees [$(\omega_{RF}+\delta\omega_{RF})t_{p}=\frac{37}{36}\pi$  instead of $\omega_{RF}t_{p}=\pi$] was studied, whereas in Fig. \ref{fig:fig2}(B), the pulses were applied with a resonance offset of $\frac{\omega_{\text{off}}}{2\pi}=300$ Hz and with $\delta\omega_{RF}=0$. As predicted from Eq. (\ref{eq:cpmgprops}), both the (green curves) CPMG($X,\,X$) and (blue curves) CPMG($Y,-Y$) were unable to spin-lock the initial $\widehat{y}$-magnetization generated from a  $\left(\frac{\pi}{2}\right)_{X}$ excitation pulse since the effective ``pulses'' generated by these sequences have phase $X$, $\widehat{R}^{\text{eff.}}_{X}$ in Eq. (\ref{eq:cpmgprops}). Therefore the echo amplitudes under the CPMG($X,\,X$) and CPMG($Y,-Y$) pulse trains died out on a timescale comparable to the decay in the FID [magenta curve] in both Figs. \ref{fig:fig2}(A) and \ref{fig:fig2}(B).  In Fig. \ref{fig:fig2}(A), the CPMG($Y,\,Y$) pulse train lead to long-lived echoes [black curve], whereas the CPMG($X,-X$) pulse train [red curve] did not generate any long-lived echoes.  As mentioned above, this was due to the dependence of the spin-locking amplitude [Eq. (\ref{eq:amplo})] in $\overline{H}^{(1)}_{\text{avg}}$ on pulse flip errors:  for transitions with $\omega_{jk}^{m,m-1}\tau\approx 0$, $\lambda_{kj}^{m,m-1}\approx 0$ and $\lambda_{kj}^{m,m-1}\neq 0$ for CPMG($X,-X$) and CPMG($Y,\,Y$) pulse trains, respectively.  This means for those transitions, the effective spin-locking field is approximately 0 Hz for the CPMG($X,-X$) pulse train and 55.6 Hz for the CPMG($Y,\,Y$) pulse train.  With nonzero resonance offsets, however, both the CPMG($X,-X$) [red curve, Fig. \ref{fig:fig2}(B)] and CPMG($Y,\,Y$) [black curve, Fig. \ref{fig:fig2}(B)] lead to long-lived echoes due to the fact that both sequences were able to generate a nonzero spin-locking field in $\overline{H}^{(1)}_{\text{avg}}$ with a strength proportional to $\left|\cos(\omega_{\text{off}}\tau)\right|=0.876$ for the CPMG($X,-X$) pulse train and $\left|\sin(\omega_{\text{off}}\tau)\right|=0.482$ for the CPMG($Y,\,Y$) pulse train.

\begin{figure}
\includegraphics*[scale=.5]{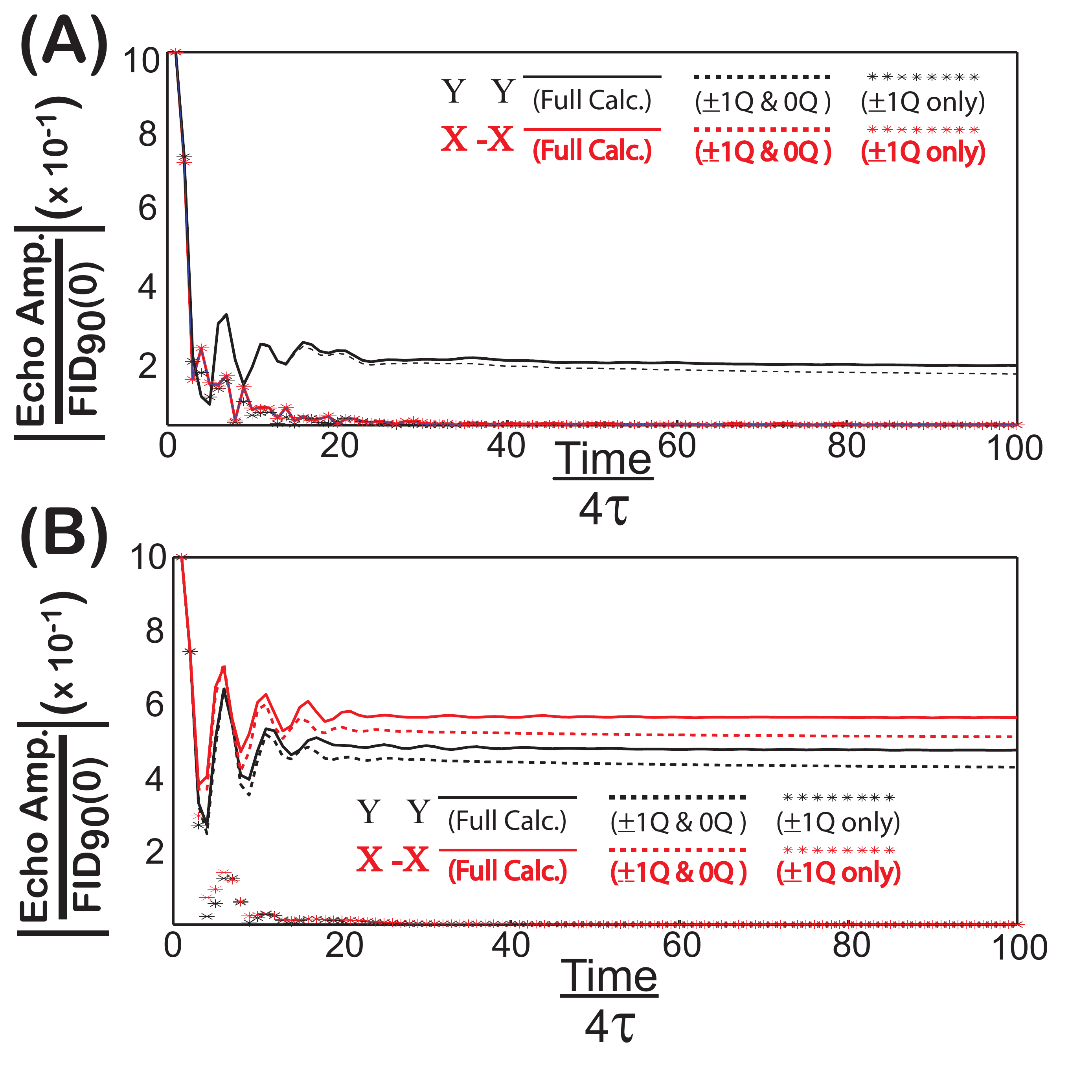}
\caption{(Color online) The effects of coherence selection during the (black curves) CPMG($Y,\,Y$) and (red curves) CPMG($X,-X$) pulse trains applied to a system of 10 spins, averaged over 200 different linear arrangements of the spins under the same conditions used in Fig. \ref{fig:fig2}.  The solid lines represent the normalized echo amplitudes without any selection (similar to the curves in Fig. \ref{fig:fig2}), the dashed lines represent the normalized echo amplitudes where both zero- and single-quantum coherences were retained after each CPMG pulse block, and the asterisks (****) represent the normalized echo amplitudes when only single-quantum coherences were retained.  As can be seen from the similarity between the echo amplitudes under no selection (solid lines) and with zero-/single-quantum coherence selection (dashed lines), the echo amplitudes are mainly governed by dynamics within the zero-quantum/single-quantum subspace.
}
\label{fig:fig2a}
\end{figure}

From the simulations in Fig. \ref{fig:fig2}, a plateau in the echo amplitudes was reached around $n_{l}=20$ under the CPMG($Y,\,Y$) [Fig. \ref{fig:fig2}(A) and Fig. \ref{fig:fig2}(B)] and CPMG($X,-X$) [Fig. \ref{fig:fig2}(B)] pulse trains.  This plateau can be thought of as arising from a periodic quasiequiibrium\cite{Sakellariou98} generated by the CPMG pulse trains that possesses a nonzero contribution corresponding to $\widehat{y}$-magnetization.  It has long been known\cite{Ivanov78,Suwelack80,Maricq85} that pulsed spin-locking of magnetization in dipolar solids generates an initial quasiequilibrium state that leads to spin-locked transverse magnetization.  While the $\widetilde{R}^{\text{eff.}}_{\Psi_{1}}$ excitation propagators in Eq. (\ref{eq:cpmgprops}) do not necessarily correspond to an RF pulse like those used in pulsed spin-locking, an analogous expression for the quasiequilibrium under the effective ``pulsed'' spin-locking of $\widehat{\rho}(0)$ in Eq. (\ref{eq:ymag}) under the CPMG propagators in Eq. (\ref{eq:cpmgprops}), $\widehat{\rho}_{QE}$, can be written as:
\begin{eqnarray}
\widehat{\rho}_{QE}&\approx&\beta_{QE}\left(\frac{2t_{p}}{2t_{p}+4\tau}\overline{H}_{\text{avg}}+\frac{4\tau}{2t_{p}+4\tau}\widehat{H}_{D}\right)=\beta_{QE}\widetilde{H}_{\text{eff}}
\label{eq:quasi}
\end{eqnarray}
where $\widetilde{H}_{\text{eff}}$ is the average Hamiltonian under a CPMG($\phi_{1},\phi_{2}$) pulse train ($n_{l}=1$) and was derived from the propagator in Eq. (\ref{eq:cpmg1loop}).  The temperature, $\beta_{QE}$, is given by:
\begin{eqnarray}
\beta_{QE}&=&\frac{\text{Trace}\left[\widetilde{H}_{\text{eff}}\widehat{\rho}(0)\right]}{\text{Trace}\left[\left(\widetilde{H}_{\text{eff}}\right)^{2}\right]}=-\frac{\text{Trace}\left[\widetilde{H}_{\text{eff}}\widehat{I}_{Y}\right]}{\text{Trace}\left[\left(\widetilde{H}_{\text{eff}}\right)^{2}\right]}\nonumber\\
\end{eqnarray}

In writing Eq. (\ref{eq:quasi}), the following assumptions, similar to the conditions needed under pulsed spin-locking\cite{Maricq85}, are necessary:  $\widetilde{H}_{\text{eff}}$ is the only constant of motion for the spin system and $\tau\neq 0$, $\left|\frac{2t_{p}}{\hbar}\overline{H}_{\text{avg}}\right|\ll \widehat{1}$ and that the spectrum of $\widehat{H}_{D}$ lies within the range $\pm \frac{\pi}{2\tau}$.  With $\widehat{\rho}_{QE}$ in Eq. (\ref{eq:quasi}), the fraction of transverse magnetization that is expected to be ``preserved'' under the CPMG($\phi_{1},\phi_{2}$) pulse train is given by:
\begin{eqnarray}
\frac{\langle \widehat{I}_{Y}\rangle_{QE}}{\langle \widehat{I}_{Y}(0)\rangle}&=&\beta_{QE}\frac{\text{Trace}\left[\widehat{I}_{Y}\widetilde{H}_{\text{eff}}\right]}{\text{Trace}\left[\widehat{I}_{Y}\widehat{\rho}(0)\right]}=\frac{\left(\text{Trace}\left[\widetilde{H}_{\text{eff}}\widehat{I}_{Y}\right]\right)^{2}}{\text{Trace}\left[\left(\widehat{I}_{Y}\right)^{2}\right]\text{Trace}\left[\left(\widetilde{H}_{\text{eff}}\right)^{2}\right]}\nonumber\\
\label{eq:qem}
\end{eqnarray}

If $\overline{H}_{\text{avg}}\approx \overline{H}_{\text{avg}}^{(1)}$, Eq. (\ref{eq:qem}) can be rewritten in a compact form as:
\begin{eqnarray}
\frac{\langle \widehat{I}_{Y}\rangle_{QE}}{\langle \widehat{I}_{Y}(0)\rangle}&\approx&\frac{\cos^{2}(\Delta\chi)\sin^{2}(\Psi_{1})\left[\text{FID}_{D}(\tau)\right]^{2}}{\frac{1+\cos(2\Delta\chi)\text{FID}_{D}(2\tau)}{2}+4\left(\frac{\omega_{loc}\tau}{\delta\omega t_{p}}\right)^{2}}
\label{eq:magfin}
\end{eqnarray}
where FID$_D(\tau)$ is the FID under pure dipolar evolution at time $\tau$:
\begin{eqnarray}
\text{FID}_{D}(\tau)&=&\frac{\text{Trace}\left[\widehat{I}_{Y}\widehat{U}_{D}(\tau)\widehat{I}_{Y}\widehat{U}^{\dagger}_{D}(\tau)\right]}{\text{Trace}\left[\left(\widehat{I}_{Y}\right)^{2}\right]}
\end{eqnarray}
and $\omega_{loc}=\sqrt{\frac{\text{Trace}\left[\left(\widehat{H}_{D}\right)^{2}\right]}{\text{Trace}\left[\left(\widehat{I}_{Y}\right)^{2}\right]}}$.

Due to the contributions of $\overline{H}^{(1)}_{\text{avg}}$ in Eq. (\ref{eq:avgH1}) to $\widehat{\rho}_{QE}$, $\widehat{y}$-magnetization will be preserved under the CPMG($Y,\,Y$) and CPMG($X,-X$) pulse trains since $\overline{H}_{\text{avg}}$ has a phase of $\Psi_{1}=\pm Y$, $\sin^{2}(\Psi_{1})=1$, and therefore $\langle \widehat{I}_{Y}\rangle_{QE}\neq 0$, whereas for the CPMG($X,\,X$) and CPMG($Y,-Y$) sequences, $\overline{H}_{\text{avg}}^{(1)}$ has a phase of $\Psi_{1}=X$, $\sin^{2}(\Psi_{1})=0$, and therefore $\langle \widehat{I}_{Y}\rangle_{QE}=0$.  In the simulations in Fig. \ref{fig:fig2}(A), the magnitude of the echo amplitude (for $n_{l}\geq 40$) relative to the initial $\widehat{y}$-magnetization was $\overline{\frac{\langle \widehat{I}_{Y}\rangle_{QE}}{\langle\widehat{I}_{Y}(0)\rangle}}=0.1549$ under the CPMG($Y,\,Y$) pulse train [calculated value using Eq. (\ref{eq:qem}) was $\overline{\frac{\langle \widehat{I}_{Y}\rangle_{QE}}{\langle\widehat{I}_{Y}(0)\rangle}}=0.122$], and $\overline{\frac{\langle \widehat{I}_{Y}\rangle_{QE}}{\langle\widehat{I}_{Y}(0)\rangle}}=0.0016$ under the CPMG($X,-X$) pulse train [calculated value using Eq. (\ref{eq:qem}) was $\overline{\frac{\langle \widehat{I}_{Y}\rangle_{QE}}{\langle\widehat{I}_{Y}(0)\rangle}}=0$].  Similarly in Fig. \ref{fig:fig2}(B), the magnitude of the echo amplitude (for $n_{l}\geq 40$) relative to the initial $\widehat{y}$-magnetization was $\overline{\frac{\langle \widehat{I}_{Y}\rangle_{QE}}{\langle\widehat{I}_{Y}(0)\rangle}}=0.468$ under the CPMG($Y,\,Y$) pulse train [calculated value using Eq. (\ref{eq:qem}) was $\overline{\frac{\langle \widehat{I}_{Y}\rangle_{QE}}{\langle\widehat{I}_{Y}(0)\rangle}}=0.456$], and $\overline{\frac{\langle \widehat{I}_{Y}\rangle_{QE}}{\langle\widehat{I}_{Y}(0)\rangle}}=0.556$ under the CPMG($X,-X$) pulse train [calculated value using Eq. (\ref{eq:qem}) was $\overline{\frac{\langle \widehat{I}_{Y}\rangle_{QE}}{\langle\widehat{I}_{Y}(0)\rangle}}=0.428$].  Possible origins for the differences between the simulated quasiequilibrium magnetization, $\overline{\frac{\langle \widehat{I}_{Y}\rangle_{QE}}{\langle\widehat{I}_{Y}(0)\rangle}}$, and that calculated from Eq. (\ref{eq:qem}) for the 10 spin system are higher-order corrections to the average Hamiltonian treatment\cite{Maricq85} that render the expression for $\widehat{\rho}_{QE}$ in Eq. (\ref{eq:quasi}) only approximate, along with other constants of motion\cite{Walls06} besides $\widetilde{H}_{\text{eff.}}$.

Further numerical evidence about establishing $\widehat{\rho}_{QE}$ in Eq. (\ref{eq:quasi}) from initial $\widehat{I}_{Y}$ magnetization under the CPMG($Y,\,Y$) and CPMG($X,-X$) pulse trains is shown in Fig. \ref{fig:fig2a}.  Since $\widehat{\rho}_{QE}$ in Eq. (\ref{eq:quasi}) is predominately made up of both single-quantum [from $\overline{H}_{\text{avg}}^{(1)}$ in Eq. (\ref{eq:Havg1})] and zero-quantum terms [from $\widehat{H}_{D}$], the majority of the spin dynamics will occur within thw zero-quantum and single-quantum subspace.  A justification of this assertion is shown in Fig. \ref{fig:fig2a}, where the results from two additional simulations are shown.  In one of the simulations, only single-quantum coherences were retained after each CPMG pulse block [Fig. \ref{fig:fig2a}, denoted by $*$], and in the other simulation, both single-quantum and zero-quantum coherences  were retained after each CPMG pulse block [Fig. \ref{fig:fig2a}, dashed curves].  For comparison, the simulations with no coherence selection are also shown [Fig. \ref{fig:fig2a}, solid curves].  If only single-quantum coherences were kept (asterisks in Fig. \ref{fig:fig2a}), the echo amplitudes decayed on the same time scale as the signal under a $\left(\frac{\pi}{2}\right)_{X}-$acquire simulation [magenta curves in Fig. \ref{fig:fig2}], which is in agreement with previous numerical simulations\cite{Li08} by the Barrett group. However, if both zero-quantum and single-quantum coherences were kept [dashed curves], the echo amplitudes were within 10$\%$ of the echo amplitudes when there was no coherence selection.  The simulations in Fig. \ref{fig:fig2a} support the idea that the dynamics within the zero-quantum/single-quantum subspace describes establishing the quasiequilibrium, $\widehat{\rho}_{QE}$, from initial $\widehat{y}$-magnetization.

\begin{figure}
\includegraphics*[scale=.4]{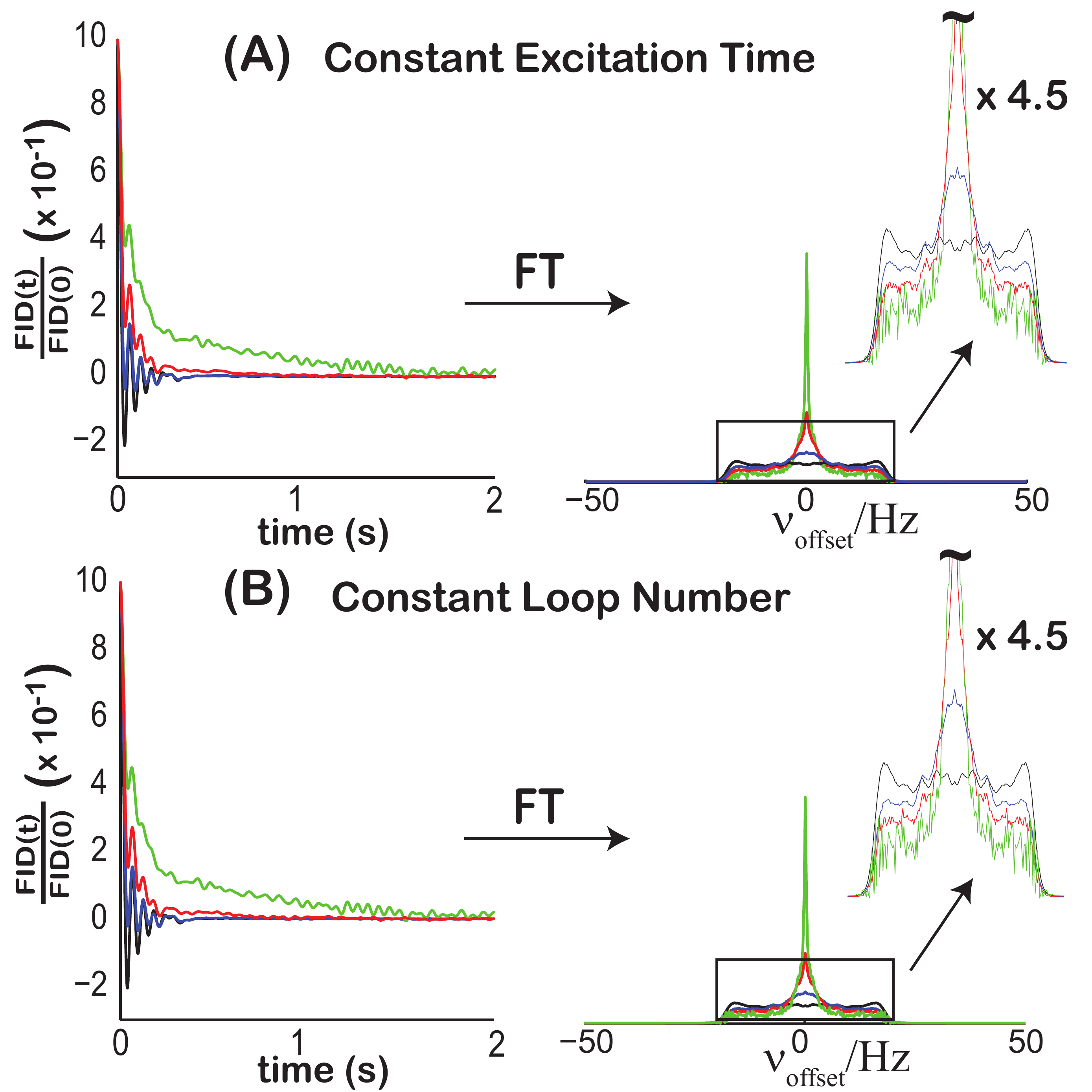}
\caption{(Color online) Simulations of the (left) normalized FIDs of the final echoes and (right) corresponding spectra after application of a $\left(\frac{\pi}{2}\right)_{X}-\left(\text{CPMG}(Y,\,Y)\right)_{n_{l}}$ sequence for (A) constant excitation time ($T_{tot}=4.64$ s) and for (B) constant CPMG loop number ($n_{l}=100$) in a system of 10 spins averaged over 1250 different linear arrangements of the spins.  The corresponding FID and spectrum from a $\left(\frac{\pi}{2}\right)_{X}$-acquire simulation are shown for comparison (black curve).  In (A), simulations for constant excitation time ($T_{tot}=4.64$ ms) are shown for the following $\tau$ and $n_{l}$ values: (blue curve) $\tau=2.9$ ms and $n_{l}=400$, (red curve) $\tau=5.8$ ms and $n_{l}$=200, and (green curve) $\tau=11.6$ ms and $n_{l}=100$.  The resulting FIDs were attenuated relative to the signal from a $\left(\frac{\pi}{2}\right)_{X}$-acquire by factors of (blue) 0.155, (red) 0.0584, and (green) 0.0265.  In (B), simulations for constant loop number ($n_{l}=100$) are shown for the following $\tau$ values:  (blue) $\tau=2.9$ ms, (red) $\tau=5.8$ ms, and (green) $\tau=11.6$ ms.  In this case, the resulting FIDs were attenuated relative to the FID from a $\left(\frac{\pi}{2}\right)_{X}$-acquire by factors of (blue) 0.1596, (red) 0.0587, and (green) 0.0265.  In both \ref{fig:fig3a}(A) and \ref{fig:fig3a}(B), increasing $\tau$ resulted in a slower decay in the FID, leading to a sharper spectrum.
}
\label{fig:fig3a}
\end{figure}

From Eq. (\ref{eq:magfin}), the echo amplitudes under the CPMG pulse trains are expected to decrease with increasing $\tau$.   This can be understood by the fact that $\overline{H}_{\text{avg}}^{(1)}$ in Eq. (\ref{eq:avgH1}) looks less like an RF pulse as $\tau$ increases, and therefore the projection of $\widetilde{H}_{\text{eff}}$ onto the initial $\widehat{y}$-magnetization decreases.  However, this decrease should depend upon the dipolar transition frequency, $\omega_{jk}^{m,m-1}$, which can be seen in the eigenstate representation of $\overline{H}_{\text{avg}}^{(1)}$ in Eq. (\ref{eq:Havg1}).  For those transitions that satisfy $\omega_{jk}^{m,m-1}\tau\ll 1$ [or in general, $|4\omega_{jk}^{m,m-1}\tau-2n\pi|\ll 1$ for some integer $n$], $\overline{H}_{\text{avg}}^{(1)}$ resembles an RF pulse, whereas for those transitions with $\omega_{jk}^{m,m-1}\tau\geq 1$, $\overline{H}^{(1)}_{\text{avg}}$ is attenuated by the transition frequency.  In this sense, coherences  with a transition frequency $\nu_{jk}^{m,m-1}\approx 0$ Hz are expected to be spin-locked and contribute most to $\widehat{\rho}_{QE}$.  A consequence of this is that the relative spectral intensity at $\nu\approx 0$ Hz should increase as $\tau$ increases.  This is clearly illustrated in Fig. \ref{fig:fig3a} which shows simulations of the FIDs [Fig. \ref{fig:fig3a}, left] and the corresponding spectra [Fig. \ref{fig:fig3a}, right] after the final echo under a CPMG($Y,\,Y$) pulse train for a 10 spin system, averaged over 1250 different linear spin arrangements, under conditions of either constant excitation time [$n_{l}\tau=4.64$ s] in Fig. \ref{fig:fig3a}(A) or constant loop number [$n_{l}=100$] in Fig. \ref{fig:fig3a}(B).  In both Fig. \ref{fig:fig3a}(A) and \ref{fig:fig3a}(B), the FID after application of the CPMG($Y,\,Y$) pulse train decayed more slowly with increasing $\tau$, leading to a narrower spectrum with a larger relative intensity at $\nu=0$ Hz.  In Fig. \ref{fig:fig3a}(B), it could be argued that the narrowing of the spectra with increasing $\tau$ was simply due to the fact that the total time of the CPMG pulse train was longer, thereby increasing the selectivity of the effective spin-locking field.  However, almost identical results were obtained in the fixed excitation time simulations ($4n_{l}\tau=4.64$ s), where once again, the longer $\tau$ (smaller $n_{l}$) simulations generated a slower decaying FID and a larger relative spectral intensity at $\nu=0$ Hz [Fig. \ref{fig:fig3a}(A)].

\section{Experimental Setup}
Experiments were performed on a 300 MHz Avance Bruker spectrometer (static magnetic field of  7T and an operating frequency for $^{1}H$
of 300.13 MHz) using a Bruker 5-mm broadband inverse (BBI) probe head at room temperature [T=298K] with RF field strengths of  $\frac{\omega_{RF}}{2\pi}=40.3$ kHz and $\frac{\omega_{RF}}{2\pi}=13$ kHz for $^{1}$H and $^{13}$C, respectively.   The ferrocene (98$\%$ purity), adamantane ($\geq99\%$ purity), $C_{60}$ (99.5$\%$ purity), and $C_{70}$ (99$\%$ purity) chemicals were all purchased from Sigma-Aldrich.  The adamantane and ferrocene samples were ground using a mortar and pestle before being placed in a 5mm NMR tube.  In all experiments, the recycle delay times were $d_{1}=25$ s for the adamantane and ferrocene samples, and $d_{1}=15$ s and $d_{1}=120$ s for the C$_{70}$ and C$_{60}$ samples, respectively.  For the adamantane and ferrocene samples, the signals under the CPMG$(\phi_{1},\phi_{2})$ and CPMG$(-\phi_{1},-\phi_{2})$ pulse trains were combined in order to remove artifacts from pulse transients. In the experiments using pulsed field gradients (PFGs) on the C$_{60}$ sample [Fig. \ref{fig:fig10}], 300$\mu$s half-sine gradient pulses were used, and a 300$\mu$s gradient stabilization delay was placed after the PFG and before echo acquisition.  In this case, the echo acquisition time between pulses, T$_{acq,E}$, was 800$\mu$s, while in all other experiments, the echo acquisition time between pulses was given by T$_{acq,E}=2\tau-10\mu$s .  All other experimental parameters [number of scans (N$_{s}$), $\tau$, and $n_{l}$] are listed in the relevant figure captions.
\section{Experimental Results and Comparison to Theoretical Predictions}
\begin{figure}
\includegraphics*[scale=.5]{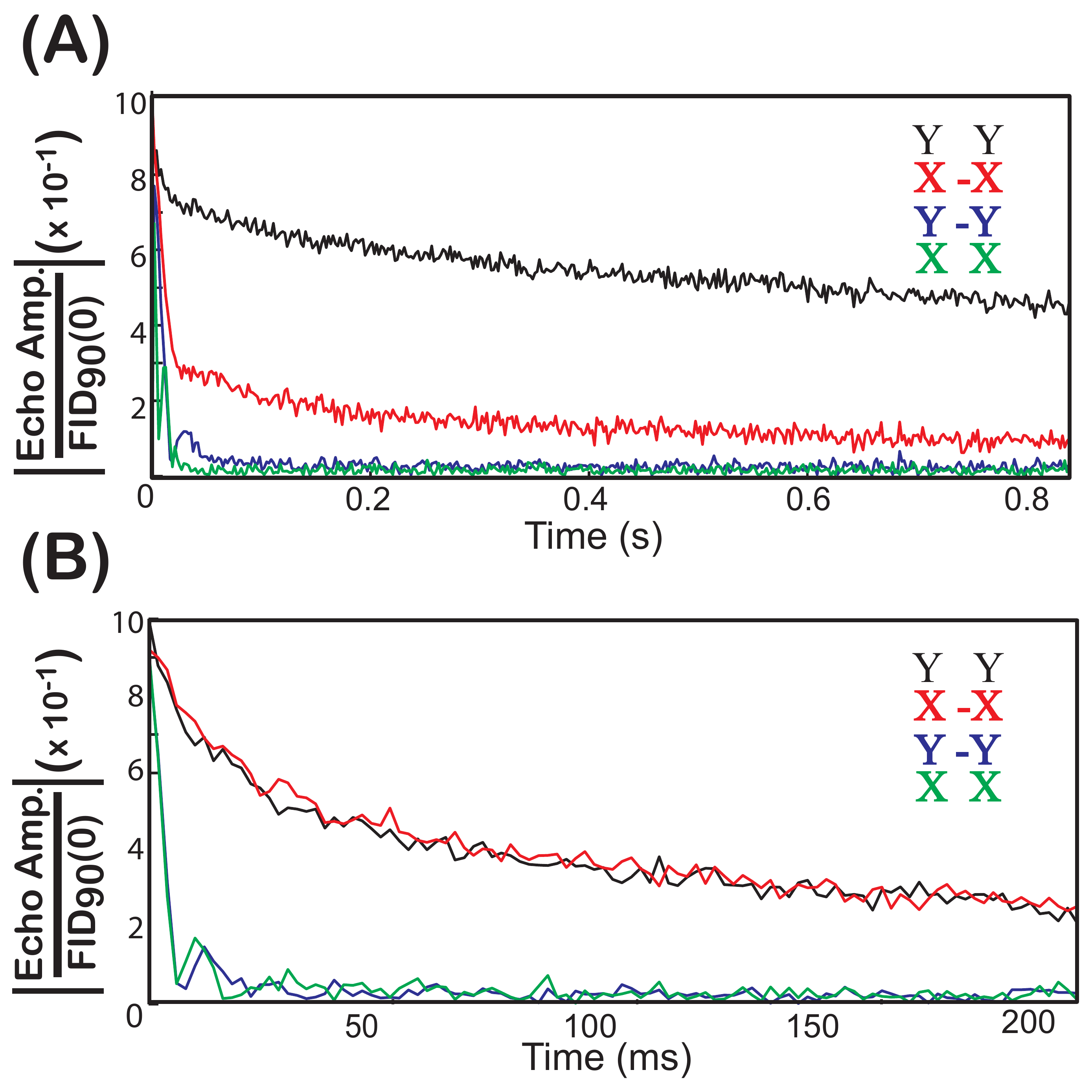}
\caption{(Color online) The echo amplitudes under the (black) CPMG($Y,\,Y$), (red) CPMG($X,-X$), (blue) CPMG($Y,-Y$), and (green) CPMG($X,\,X$) pulse trains in (A) polycrystalline C$_{60}$ [N$_s$=128, $\tau=0.5$ ms, and up to $n_{l}=400$ echo amplitudes] and (B) C$_{70}$ [N$_s$=384, $\tau=0.5$ ms and up to $n_{l}=100$ echo amplitudes] acquired on a 300.1 MHz Bruker spectrometer.  The echo amplitudes, normalized by the signal from a $\left(\frac{\pi}{2}\right)_{X}-$acquire experiment, are plotted at multiples of $4\tau$.  The CPMG($Y,\,Y$) and CPMG($X,-X$) pulse trains generated long-lived echoes for both [\ref{fig:fig3}(A)] C$_{60}$ and [\ref{fig:fig3}(B)] C$_{70}$, whereas the CPMG($X,X$) and CPMG($Y,-Y$) pulse trains did not.
}
\label{fig:fig3}
\end{figure}

\begin{figure}
\includegraphics*[scale=.5]{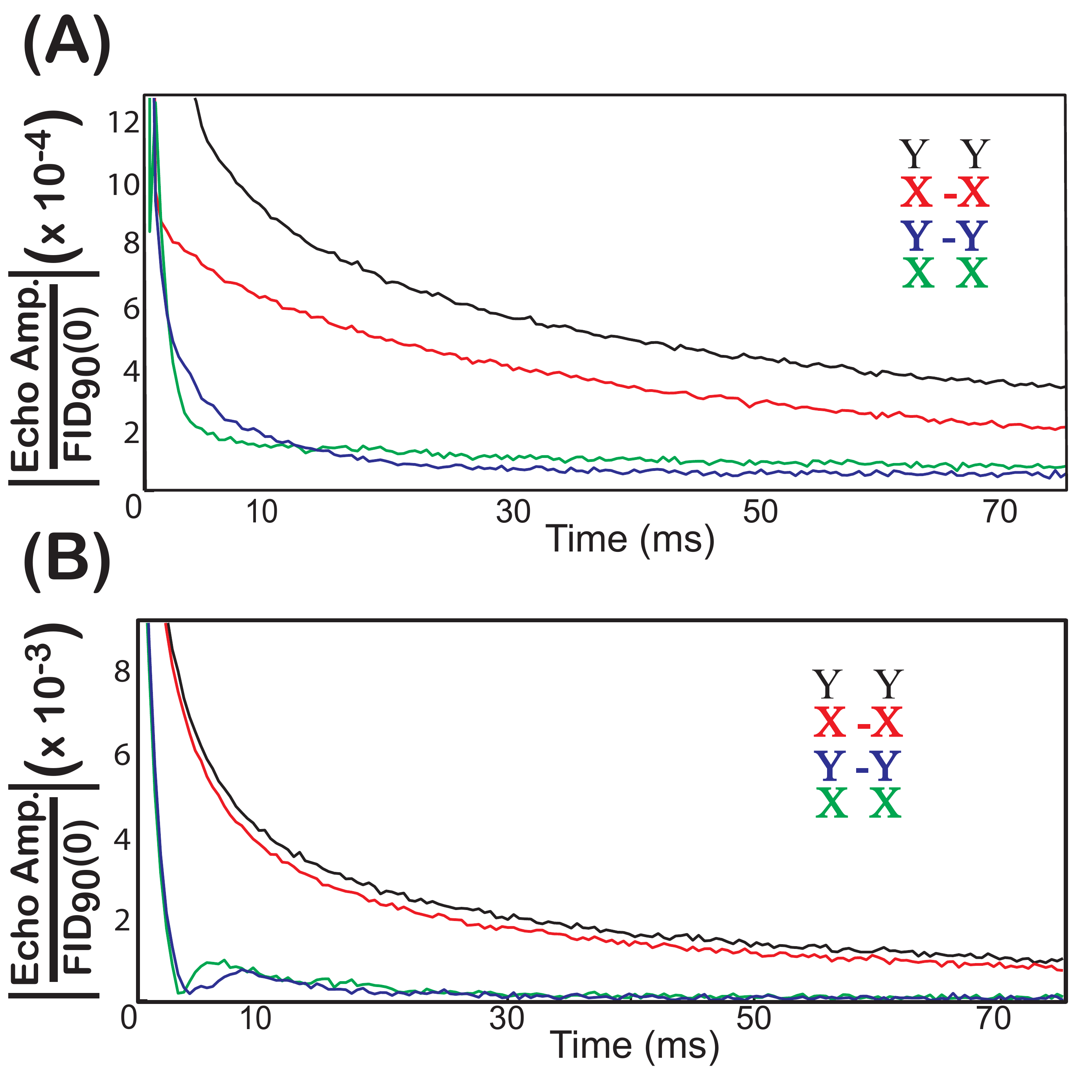}
\caption{(Color online) The echo amplitudes under the (black) CPMG($Y,\,Y$), (red) CPMG($X,-X$), (blue) CPMG($Y,-Y$), and (green) CPMG($X,\,X$) pulse trains in (A) polycrystalline adamantane [N$_s$=256, $\tau=116\mu$s and up to $n_{l}=160$ echo amplitudes] and (B) ferrocene [N$_s$=256, $\tau=116\mu$s and up to $n_{l}=160$ echo amplitudes] acquired on a 300.1 MHz Bruker spectrometer. The echo amplitudes, normalized by the signal from a $\left(\frac{\pi}{2}\right)_{X}-$acquire experiment, are plotted at multiples of $4\tau$.  The CPMG($Y,\,Y$) and CPMG($X,-X$) pulse trains generated long-lived echoes for both [\ref{fig:fig4}(A)] adamantane and [\ref{fig:fig4}(B)] ferrocene, whereas the CPMG($X,X$) and CPMG($Y,-Y$) pulse trains did not.
}
\label{fig:fig4}
\end{figure}

Two types of spin systems, magnetically dilute were investigated in this work: random spin networks and non-dilute spin networks.  In the case of the magnetically dilute spin sytems, $^{13}$C NMR was used to study the spin dynamics under the CPMG($\phi_{1},\phi_{2}$) pulse trains in both C$_{60}$ and C$_{70}$ ($^{13}$C has a natural abundance of around 1$\%$).  In C$_{60}$, all $^{13}$C atoms are chemically identical, and the C$_{60}$ molecule undergoes isotropic tumbling at room temperature that averages out chemical shift anisotropy and all intramolecular dipolar couplings in C$_{60}$ molecules containing more than one $^{13}$C atoms [approximately 12.1$\%$ of all C$_{60}$ molecules].  In C$_{70}$, there are five distinct $^{13}$C positions with different resonance frequencies;  furthermore, C$_{70}$ does not undergo isotropic tumbling at room temperature so that both chemical shift anisotropy and intramolecular dipolar couplings are not averaged away in C$_{70}$ molecules with more than one $^{13}$C atom [around $15.5\%$ of all C$_{70}$ molecules].  As a result, the spectral width in C$_{70}$ is much larger than that found in C$_{60}$.  In C$_{60}$ and C$_{70}$, the echo amplitudes under the CPMG($Y,\,Y$) and CPMG($X,-X$) pulse trains were on the same order of magnitude as the signal from a $\left(\frac{\pi}{2}\right)-$acquire experiment [Fig. \ref{fig:fig3}].  Non-dilute $^{1}$H spin networks in adamantane and ferrocene were also studied.  Adamantane is a plastic crystal with two chemically nonequivalent $^{1}$H spins that undergoes isotropic molecular tumbling at room temperature, which averages away all intramolecular dipolar coupling and chemical shift anisotropy.  In ferrocene, however, only one type of $^{1}$H spin is present, and ferrocene does not undergo isotropic tumbling at room temperature so that both chemical anisotropy and intramolecular dipolar couplings are present.  In both adamantane and ferrocene, the echo amplitudes under the CPMG($Y,\,Y$) and CPMG($X,-X$) pulse trains were many orders of magnitude smaller than the signal from a $\left(\frac{\pi}{2}\right)-$acquire experiment [Fig. \ref{fig:fig4}], which was due to the relatively strong dipole-dipole couplings in these systems and the $\tau$ values that were studied in this work, both of which result in $|\text{FID}_{D}(\tau)|\ll 1$, leading to small echo amplitudes as predicted by Eq. (\ref{eq:magfin}).

In Figure \ref{fig:fig3} and Figure \ref{fig:fig4}, the echo amplitudes under [black curve] CPMG($Y,\,Y$), [red curve] CPMG($X,-X$), [blue curve] CPMG($Y,-Y$), and [green curve] CPMG($X,\,X$) pulse trains are shown for [Fig. \ref{fig:fig3}] C$_{60}$ and C$_{70}$  and for [Fig. \ref{fig:fig4}] adamantane and ferrocene.  Although nominal $\pi-$pulses were applied, we estimated that the actual pulse flip errors were on the order of $\pm (3^{\circ}-4^{\circ})$ for the $^{13}$C experiments and $\pm (7^{\circ}-8^{
\circ})$ for the $^{1}$H experiments.  In all spin systems studied in this work, the CPMG($Y,\,Y$) and CPMG($X,-X$) pulse trains generated echo amplitudes that decayed more slowly than the echo amplitudes under the CPMG($X,\,X$) and CPMG($Y,-Y$) pulse trains, as predicted in {\bf{Section II}}.  Note that in the adamantane and ferrocene systems [Fig. \ref{fig:fig4}], there was a fast initial decay of the echo amplitudes due to strong dipole-dipole coupling, which was followed by a slower decay of a small component ($1\%$ of the initial magnetization in ferrocene and $0.1\%$ of the initial magnetization in adamantane).

\begin{figure}
\includegraphics*[scale=.5]{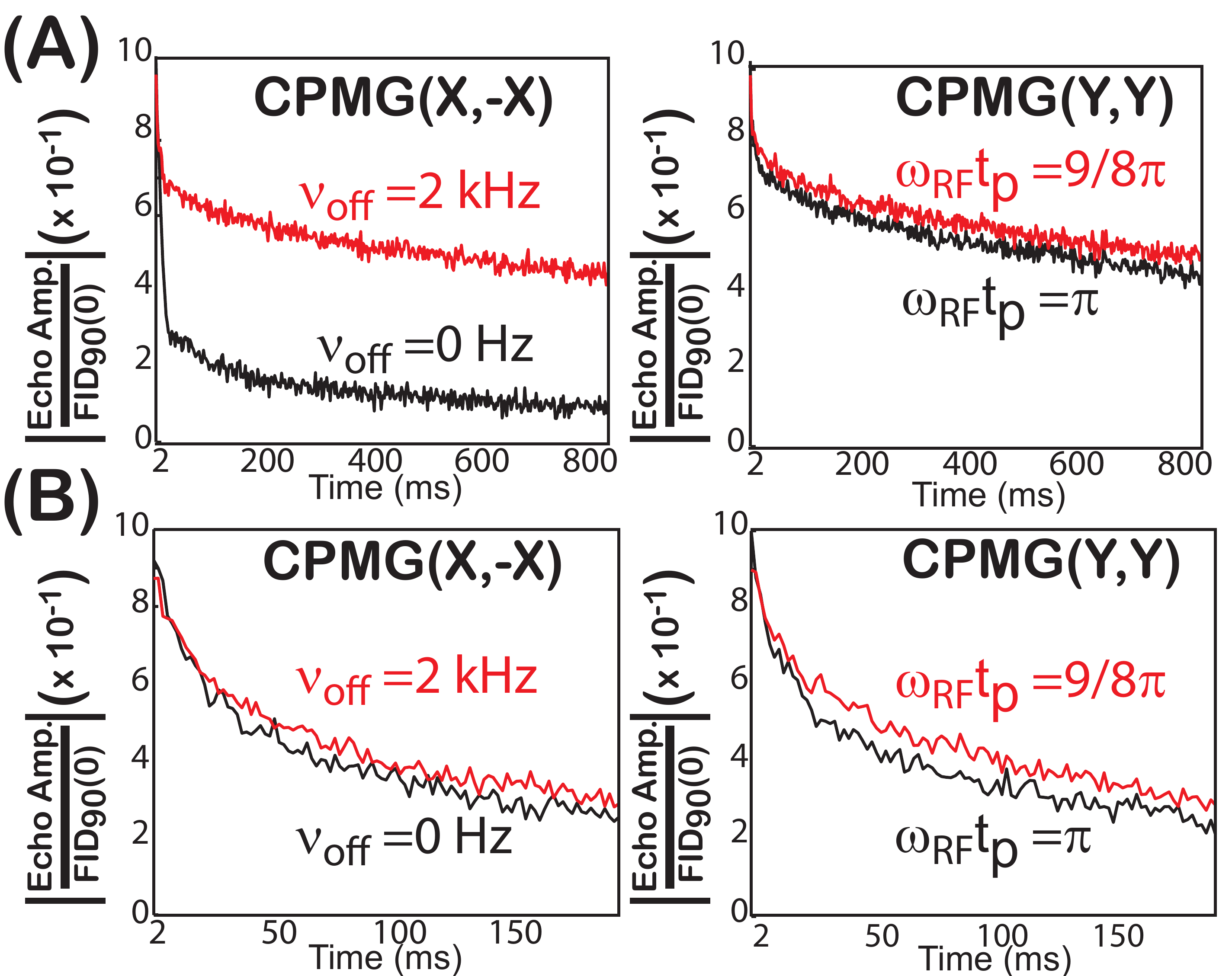}
\caption{(Color online) The effects of both (right) flip-angle errors [$\omega_{RF}t_{p}\approx \pi$ (black curve) vs. $(\omega_{RF}+\delta\omega_{RF})t_{p}\approx \frac{9}{8}\pi$ (red curve)] on the echo amplitudes under a CPMG($Y,\,Y$) pulse train and (left) resonance offsets [$\nu_{\text{off}}=0$ Hz (black curve) vs. $\nu_{\text{off}}=2$ kHz (red curve)] on the echo amplitudes under a CPMG($X,-X$) pulse train in  (A) polycrystalline C$_{60}$ [N$_s$=128, $\tau=0.5$ms] and (B) C$_{70}$ [N$_s$=384 and $\tau=0.5$ms] acquired on a 300.1 MHz Bruker spectrometer.  The echo amplitudes, which were normalized by the signal from a $\left(\frac{\pi}{2}\right)_{X}-$acquire experiment, are plotted at multiples of $4\tau$.  Pulse flip-angles did not have a significant effect on the echo amplitudes under the CPMG($Y,\,Y$) pulse trains in both C$_{60}$ and C$_{70}$ [Fig. \ref{fig:fig5}, right], indicating that the ``nominal'' $\pi-$pulses were already imperfect, i.e., $\delta\omega_{RF}\neq 0$.  However, resonance offsets generated larger echo amplitudes in C$_{60}$ [Fig. \ref{fig:fig5}(A), left] under the CPMG($X,-X$) sequence due to the relatively small chemical shift dispersion in C$_{60}$, whereas offsets did not significantly affect the echo amplitudes in C$_{70}$ [Fig. \ref{fig:fig5}(B), left] due to the already large chemical shift dispersion in C$_{70}$.
}
\label{fig:fig5}
\end{figure}

Figure \ref{fig:fig5} illustrates the effects of [left] resonance offset and [right] pulse flip-angle errors on the CPMG($X,-X$) and CPMG($Y,\,Y$) pulse trains respectively in [Fig. \ref{fig:fig5}(A)] C$_{60}$ and [Fig. \ref{fig:fig5}(B)] C$_{70}$.  For C$_{60}$ [Fig. \ref{fig:fig5}(A)], a resonance offset of $\nu_{\text{offset}}=2$ kHz [$\frac{\omega_{RF}}{2\pi}=13$ kHz] increased the echo amplitudes for the CPMG($X,-X$) sequence quite dramatically [Fig. \ref{fig:fig5}(A), left].  A similar increase in the echo amplitudes under a CPMG($X,-X$) pulse train with resonance offset was seen in the simulations shown in Fig. \ref{fig:fig2} (red curves).  Furthermore, there was a slight increase in the echo amplitude when the $\pi-$pulses were purposefully miscalibrated (202.5$^{\circ}$ pulse vs. a 180$^{\circ}$ pulse), although the increase was relatively small.  This indicated that the $\pi-$pulses were already imperfect enough to generate long-lived echoes.  In C$_{70}$ [Fig. \ref{fig:fig5}(B)], an offset of $\nu_{\text{offset}}=2$ kHz did not significantly increase the echo amplitudes [Fig. \ref{fig:fig5}(B), left], which was likely due to the fact that there exists a large chemical shift dispersion in C$_{70}$ due to the presence of nonequivalent $^{13}$C atoms and the absence of isotropic molecular tumbling at room temperature.  Again, a small increase in the echo amplitudes under a CPMG($Y,\,Y$) pulse train was observed when using miscalibrated $\pi-$pulses [Fig. \ref{fig:fig5}(B), right].

\begin{figure}
\includegraphics*[scale=.4]{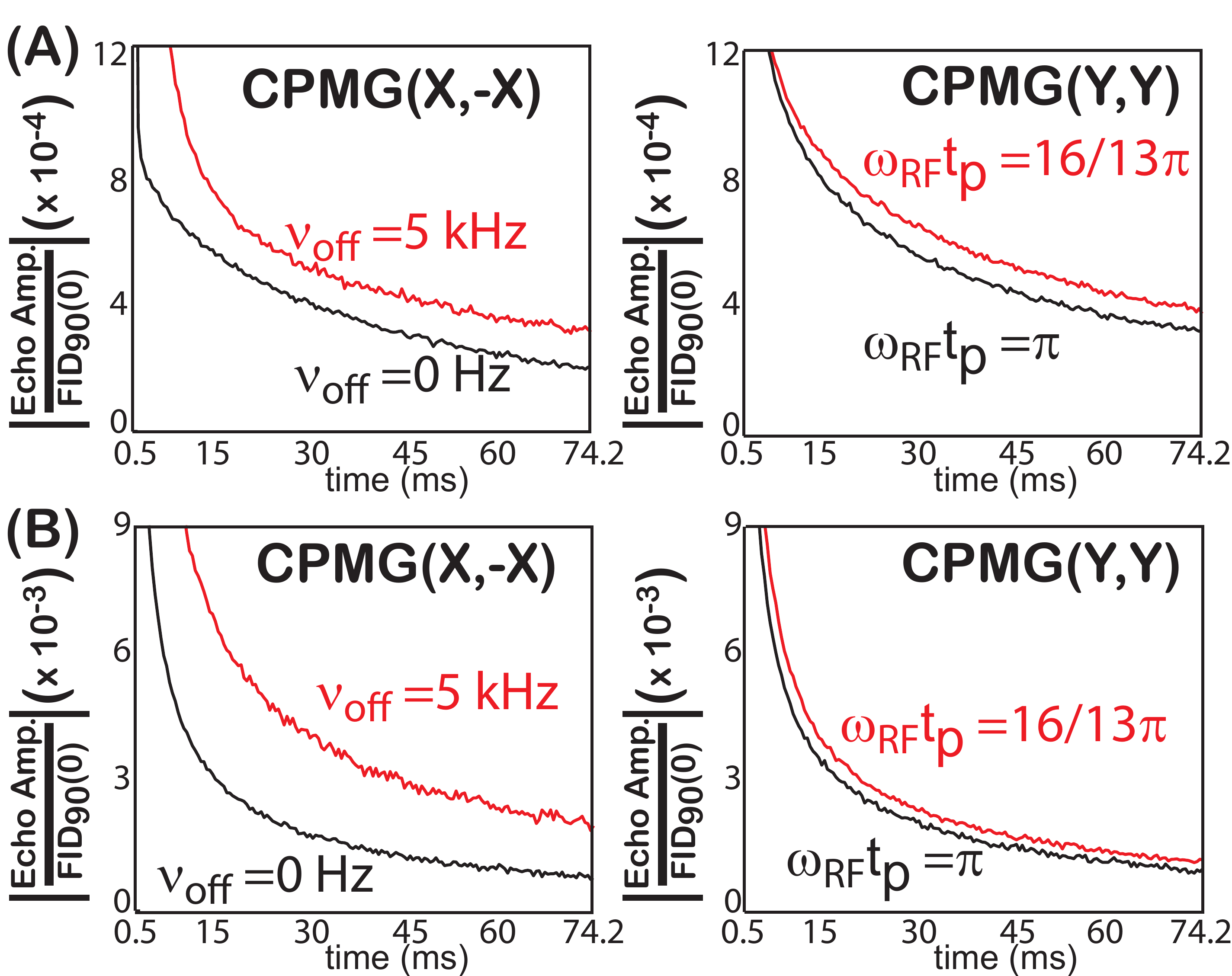}
\caption{(Color online) The effects of both (right) flip-angle errors [$\omega_{RF}t_{p}\approx \pi$ (black curve) vs. $(\omega_{RF}+\delta\omega_{RF})t_{p}\approx\frac{16}{13}\pi$ (red curve)] on the echo amplitudes under a CPMG($Y,\,Y$) pulse train and (left) resonance offsets [$\nu_{\text{off}}=0$ Hz (black curve) vs. $\nu_{\text{off}}=5$ kHz (red curve)] on the echo amplitudes under a CPMG($X,-X$) pulse train in (A) adamantane [N$_s$=256, $\tau=116\,\mu$s] and (B) ferrocene [N$_s$=256 and $\tau=116\,\mu$s] acquired on a 300.1 MHz Bruker spectrometer.  The echo amplitudes, which were normalized by the signal from a $\left(\frac{\pi}{2}\right)_{X}-$acquire experiment, are plotted at multiples of $4\tau$.  Pulse flip-angles did not have a significant effect on the echo amplitudes under the CPMG($Y,\,Y$) pulse trains in both admantane and ferrocene [Fig. \ref{fig:fig6}, right], indicating that the ``nominal'' $\pi-$pulses were already imperfect, i.e., $\delta\omega_{RF}\neq 0$.  However, resonance offsets generated larger echo amplitudes in both adamantane [Fig. \ref{fig:fig6}(A), left] and ferrocene [Fig. \ref{fig:fig6}(B), left] under the CPMG($X,-X$) pulse train, which was a consequence of the small chemical shift dispersion in these samples.
}
\label{fig:fig6}
\end{figure}

Figure \ref{fig:fig6} illustrates the effects of [left] resonance offset and [right] pulse flip-angle errors on the CPMG($X,-X$) and CPMG($Y,\,Y$) pulse trains, respectively, in [Fig. \ref{fig:fig6}(A)] adamantane and [Fig. \ref{fig:fig6}(B)] ferrocene.  Application of $\pi-$pulses applied with a resonance offset of $\nu_{\text{offset}}=5$ kHz [$\frac{\omega_{RF}}{2\pi}=40.3$ kHz] lead to an increase in the echo amplitudes under a CPMG($X,-X$) pulse train in both adamantane [Fig. \ref{fig:fig6}(A), left] and ferrocene [Fig. \ref{fig:fig6}(B), left], similar to the results in C$_{60}$ [Fig. \ref{fig:fig5}(A), left].  This increase in the echo amplitudes was consistent with the relatively small chemical shift dispersion in both adamantane and ferrocene, which is often the case for $^{1}$H spins compared with $^{13}$C spins.  Similar to what was seen in both C$_{60}$ and C$_{70}$ [Fig. \ref{fig:fig5}, right], there was a slight increase in the echo amplitudes when the $\pi-$pulses were purposefully miscalibrated (221.5$^{\circ}$ pulses vs. 180$^{\circ}$ pulses), although the increase was relatively small.  This again indicated that the $\pi-$pulses were already imperfect (flip-error of approximately $\pm 7^{\circ}$).

\begin{figure}
\includegraphics*[scale=.4]{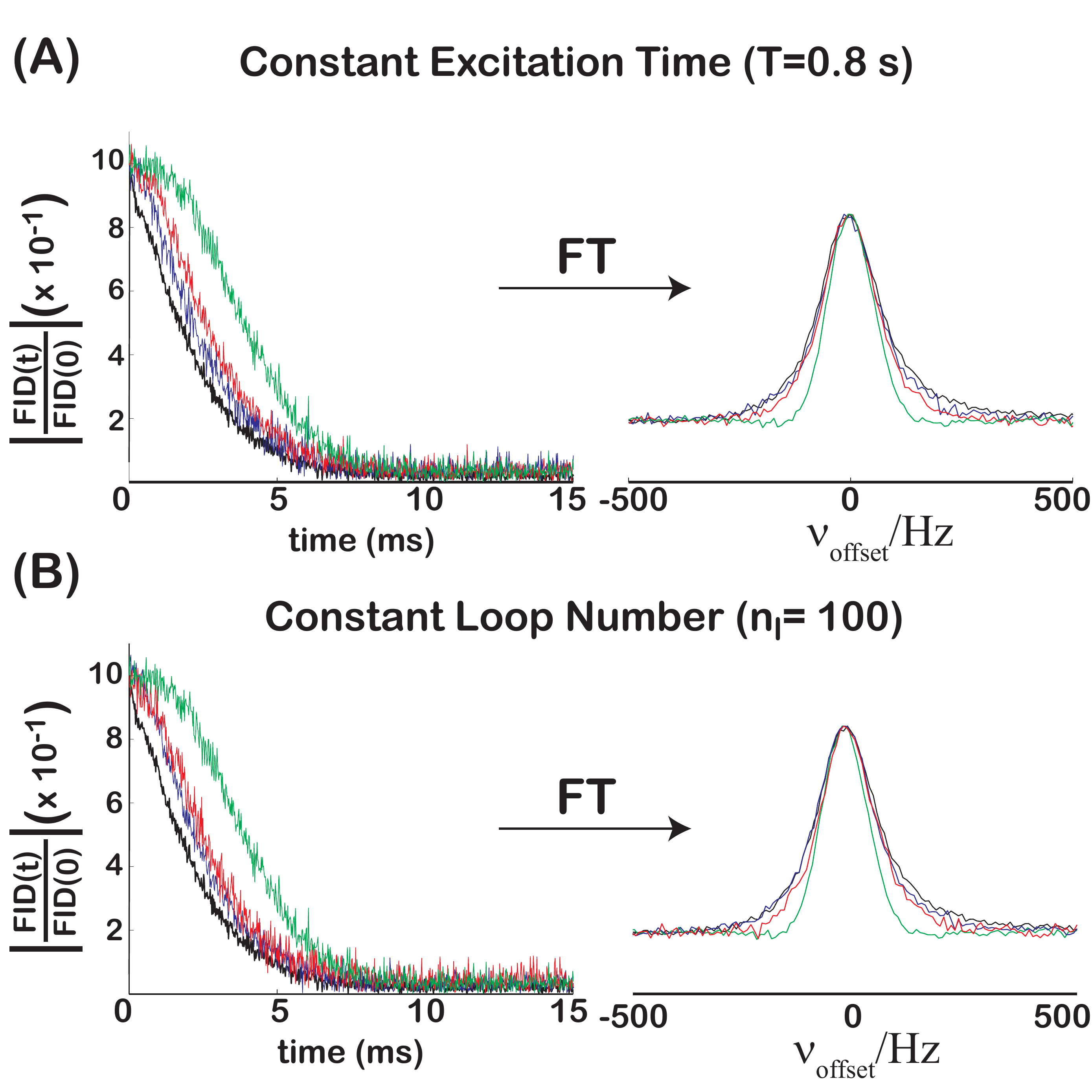}
\caption{(Color online) The absolute value of the (left) FIDs for the last echo, normalized by the echo amplitude, and (right) the corresponding spectra generated from a $\left(\frac{\pi}{2}\right)_{X}-\left(\text{CPMG}(Y,Y)\right)_{n_{l}}$ sequence under conditions of (A) constant excitation time, $T_{tot}=800$ ms, and (B) constant CPMG loop number, $n_{l}=100$, in a polycrystalline C$_{60}$ sample acquired on a 300.1 MHz Bruker spectrometer.  The FID and spectrum from a $\left(\frac{\pi}{2}\right)_{X}$-acquire experiment are shown for comparison (black curve, N$_s$=128).  (A) For the constant excitation time experiments, the following $n_{l}$, $\tau$, and N$_{s}$  values were used: (blue curve) $\tau=0.5$ ms, $n_{l}=400$ and N$_s$=128, (red curve) $\tau=1$ ms, $n_{l}$=200, and N$_s$=128, and (green curve) $\tau=2$ ms, $n_{l}=100$, and N$_s=$1024.  The corresponding echo amplitudes were attenuated (relative to the signal from a $\left(\frac{\pi}{2}\right)_{X}$-acquire experiment) by factors of (blue) 0.46, (red) 0.33, and (green) 0.15.  (B) For the constant $n_{l}=100$ experiments, the following $\tau$ and N$_{s}$ values were used:  (blue) $\tau=0.5$ ms with N$_s$=128, (red) $\tau=1$ ms with N$_s=$256, and (green) $\tau=2$ ms with N$_s=$1024.  The corresponding echo amplitudes were attenuated (relative to the signal from a $\left(\frac{\pi}{2}\right)_{X}$-acquire experiment) by factors of (blue) 0.63, (red) 0.43, and (green) 0.15.  In both (A) and (B), as $\tau$ increased, the FID decayed more slowly, resulting in a sharper spectrum.
}
\label{fig:fig7}
\end{figure}

\begin{figure}
\includegraphics*[scale=.4]{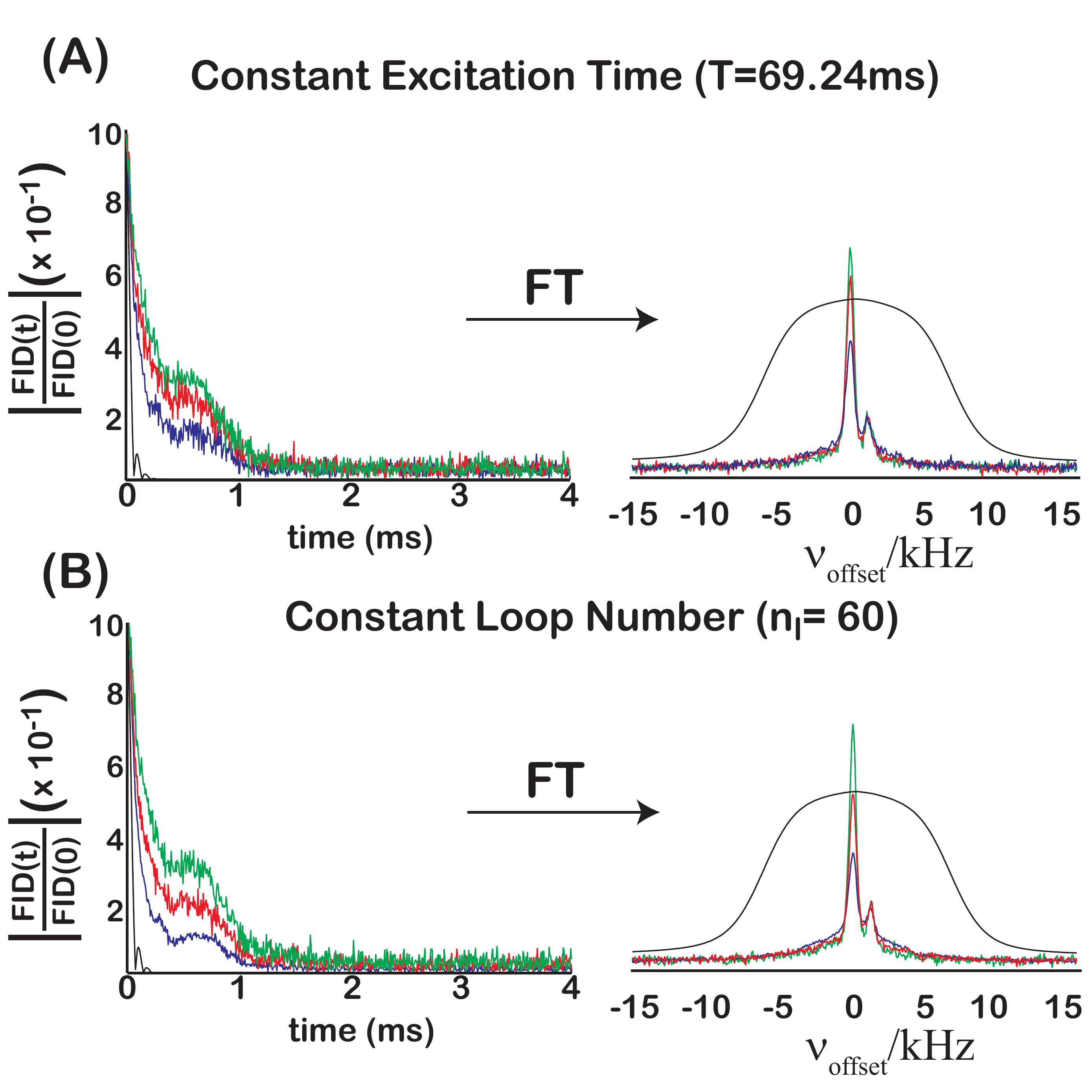}
\caption{(Color online) The absolute value of (left) the FID for the last echo, normalized by the echo amplitude, and (right) the corresponding spectra generated by averaging the  $\left(\frac{\pi}{2}\right)_{X}-\left(\text{CPMG}(Y,\,Y)\right)_{n_{l}}$ and $\left(\frac{\pi}{2}\right)_{X}-\left(\text{CPMG}(-Y,-Y)\right)_{n_{l}}$ sequences under conditions of  (A) constant excitation time, $T_{tot}\approx 69.24$ ms, and (B) constant CPMG loop number, $n_{l}=60$, in a polycrystalline adamantane sample acquired on a 300.1 MHz Bruker spectrometer.  The FID from a $\left(\frac{\pi}{2}\right)_{X}$-acquire experiment and spectrum are shown for comparison (black curve, N$_s$=128).  (A) For the constant excitation time experiments,  the following $n_l$, $\tau$, and N$_{s}$ values were used: (blue curve) $\tau=0.117$ ms, $n_{l}=148$, and N$_s=$256, (red curve) $\tau=0.216$ ms, $n_{l}$=80, and N$_s=$512, and (green curve) $\tau=0.321$ ms, $n_{l}=54$, and N$_s$=1024.  The corresponding echo amplitudes were attenuated (relative to the signal from a $\left(\frac{\pi}{2}\right)_{X}$-acquire experiment) by factors of (blue) $2.7\times 10^{-4}$, (red) $1.6\times 10^{-4}$, and (green) $1.2\times 10^{-4}$.  (B) For the constant $n_{l}=60$ experiments, the following $\tau$ and N$_{s}$ values were used: (blue) $\tau=0.116$ ms with N$_s=512$, (red) $\tau=0.216$ ms with N$_s=$512, and (green) $\tau=0.316$ ms with N$_s$=1024.  The corresponding echo amplitudes were attenuated (relative to the signal from a $\left(\frac{\pi}{2}\right)_{X}$-acquire experiment) by factors of (blue) $4.8\times 10^{-4}$, (red) $2.0\times 10^{-4}$, and (green) $1.1\times 10^{-4}$. In both (A) and (B), as $\tau$ increased, the FID decayed more slowly, resulting in a sharper spectrum.
}
\label{fig:fig8}
\end{figure}

The effects of a CPMG($Y,\,Y$) pulse train on the FID and corresponding spectrum of the final echo were investigated in both [Fig. \ref{fig:fig7}] C$_{60}$ and in [Fig. \ref{fig:fig8}] adamantane.  In both C$_{60}$ and adamantane, increasing $\tau$ lead to a slower decay of the FID, with the largest $\tau$ values decaying the slowest [green curves in Figs. \ref{fig:fig7}-\ref{fig:fig8}].  This was observed under conditions of both constant total excitation time [Fig. \ref{fig:fig7}(A) for C$_{60}$ and Fig. \ref{fig:fig8}(A) for adamantane] and for constant loop number [Fig. \ref{fig:fig7}(B) for C$_{60}$ and Fig. \ref{fig:fig8}(B) for adamantane].  The CPMG($Y,\,Y$) pulse train with the largest $\tau$ value led to the sharpest spectra in all cases, which was similar to the simulation results in Fig. \ref{fig:fig3a}.  As discussed in {\bf{Section II}}, those transitions with $\omega_{jk}^{m,m-1}\tau\ll 1$ contribute most to the echo amplitudes under the CPMG($Y,\,Y$) pulse trains, thus explaining the sharpening in the spectra around $\nu\approx 0$ Hz.  One interesting note is the two resonances that appear in the adamantane spectra in Fig. \ref{fig:fig8}.  While it might be tempting to attribute the two peaks to the two different types of $^{1}$H spins in an adamantane molecule (similar experiments on ferrocene contained only a single peak in the spectra[data not shown]), the frequency difference between the two peaks [$\Delta\delta\approx 4.5$ ppm] was much larger than the isotropic chemical shift difference between the two $^{1}$H spins found in solution [$\Delta\delta\approx 0.12$ ppm for adamantane dissolved in deuterated chloroform].  The fine structure could be the result of some sort of orientation selection under the CPMG($Y,\,Y$) pulse train\cite{Logan02,Pell11} or possibly a consequence of exponentially damped, sinusoidal modulations in the FID recently observed in dipolar soilds\cite{Morgan08}.  Further experimental work needs to be performed in order to characterize these spectral features in adamantane.

 The experimental results in Figs. \ref{fig:fig3}-\ref{fig:fig8} have all been interpreted through the prism of spin thermodynamics and average Hamiltonian theory, where the CPMG($Y,\,Y$) and CPMG($X,-X$) pulse trains, whose propagators [Eq. (\ref{eq:cpmgprops})] were shown to be similar to the propagators under pulsed spin-locking of $\widehat{y}$-magnetization, generate a quasiequilibrium with nonzero $\widehat{y}$-magnetization.  The spin dynamics under a CPMG($\phi_{1},\phi_{2}$) pulse train can also be viewed in terms of coherence transfer pathways\cite{Bain84,Bodenhausen84}.  It is well known that many coherence transfer pathways are generated from a CPMG pulse train with imperfect $\pi-$pulses.  Even for a single spin-1/2,  there are potentially $3^{n_{l}}$ coherence transfer pathways generated from a CPMG pulse train.  The periodicity of the CPMG pulse train also ensures that all possible coherence pathways that begin and end with single-quantum coherence will refocus at the time of the echo, meaning that the long-lived echoes contain contributions from all possible coherence pathways\cite{Goelman95,Hung10}.  As mentioned in the introduction, one proposal\cite{Franzoni05,Franzoni08,Levstein08,Franzoni12} for the long-lived echoes under the CPMG pulse train is that the echoes arise from stimulated echo\cite{Hahn50} pathways.  In a stimulated echo pathway, a portion of the single-quantum coherence is stored as zero-quantum coherence and/or longitudinal magnetization by an imperfect $\pi-$pulse, which can then be transferred back into single-quantum coherence from subsequent imperfect $\pi-$pulses.  Due to the fact that the longitudinal relaxation time is often longer than the transverse relaxation time, i.e., T$_{1}> $T$_{2}$, pathways where the coherence is stored as longitudinal magnetization during the CPMG pulse train will, in general, decay at a slower rate than coherence pathways where the magnetization is never stored along the $\widehat{z}-$axis.

\begin{figure}
\includegraphics*[scale=.3]{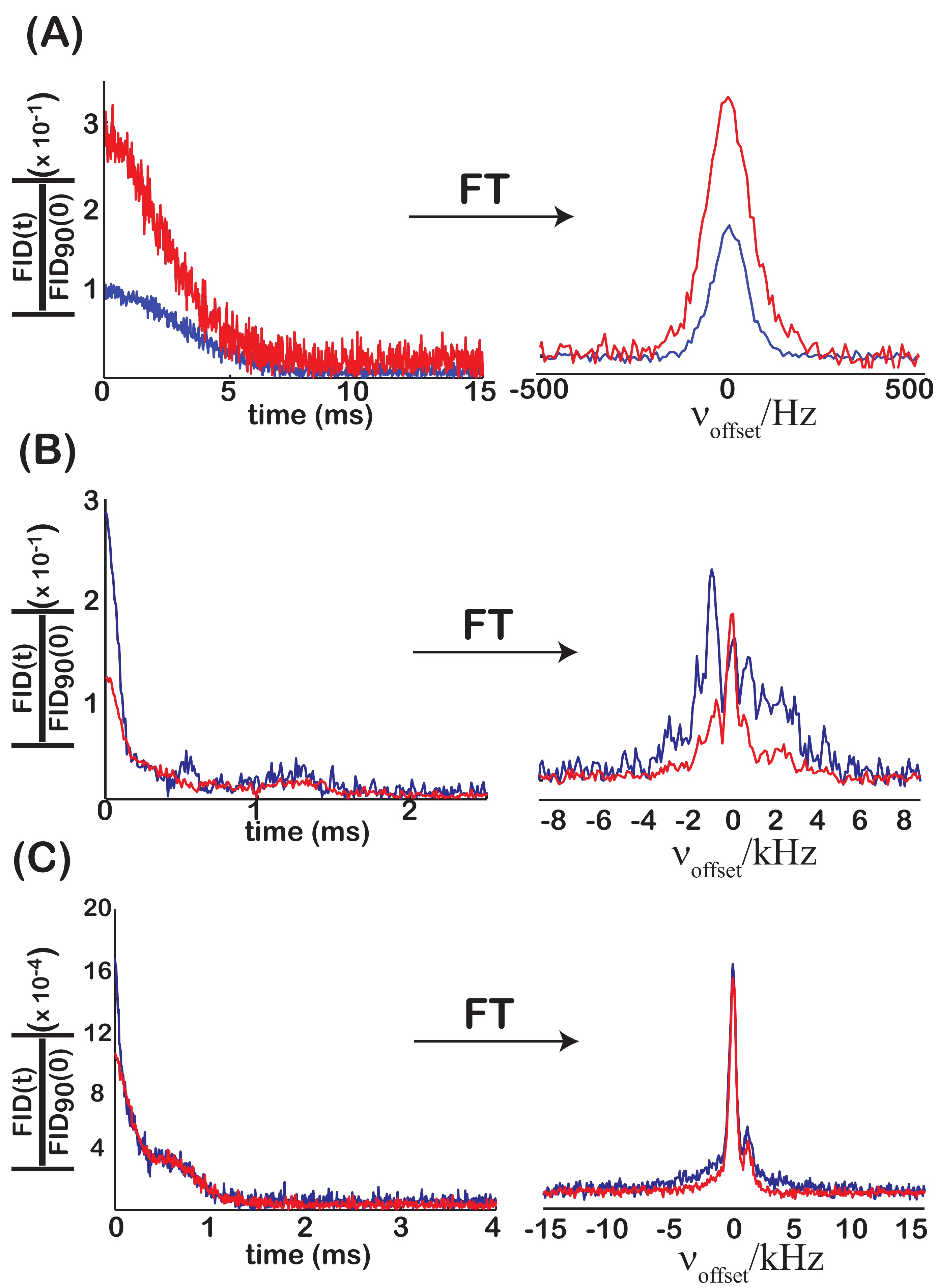}
\caption{(Color online) The absolute value of (left) the FIDs, which were normalized by the signal from a $\left(\frac{\pi}{2}\right)_{X}-$acquire experiment, and (right) the spectra generated from a randomized CPMG($Y,\,Y$) pulse train in Eq. (\ref{eq:rando}), and averaged over (A) 10 sets of random delays in C$_{60}$ [$T_{tot}=0.48$ s, $n_{l}=100$, $\tau_{k}\in\left[0.4\,\text{ms},\,2\,\text{ms}\right]$ with $\overline{\langle \tau\rangle} =1.2$ ms, and N$_s=32$ for each set of random delays], (B) 20 sets of random delays in C$_{70}$ [$T_{tot}=58.8$ ms, $n_{l}=20$, $\tau_{k}\in\left[0.22\,\text{ms},\,1.2\,\text{ms}\right]$ with $\overline{\langle \tau\rangle}=0.735$ ms, and N$_s=512$ for each set of random delays], and (C) 10 sets of random delays in adamantane $[T_{tot}=69.84$ ms, $n_{l}=90$, $\tau_{k}\in\left[80\,\mu\text{s},\,320\,\mu\text{s}\right]$ with $\overline{\langle \tau\rangle}=194\,\mu$s, and N$_s=128$ for each set of random delays].  The corresponding constant $\tau$ experiments (red curves) are also shown for comparison: in (A) C$_{60}$, $n_{l}=100$, $\tau=1.2$ ms, $T_{tot}=4n_{l}\tau=0.48$ s, and N$_s=45$, in (B) C$_{70}$, $n_{l}=20$  $\tau=0.735$ ms, $T_{tot }=58.8$ms, and N$_s$=2048, and in (C) adamantane, $n_{l}=90$, $\tau=194\mu$s, $T_{tot}=69.84$ms, and N$_s=512$.  The spectra were narrower from the randomized CPMG($Y,\,Y$) compared with the spectra from the regular CPMG($Y,\,Y$).
}
\label{fig:fig9}
\end{figure}

 In order to investigate the contributions of various coherence transfer pathways to the long-lived echoes during a CPMG($Y,\,Y$) pulse train, the periodicity of the CPMG($Y,\,Y$) pulse train was purposely broken by using randomly chosen delays as follows:
 \begin{eqnarray}
 \left(\frac{\pi}{2}\right)_{X}-\prod^{n=n_{l}}_{n=1}\left(\tau_{n}-(\pi)_{Y}-2\tau_{n}-(\pi)_{Y}-\tau_{n}\right)
 \label{eq:rando}
  \end{eqnarray}
  where $\tau_{k}\neq\tau_{j}$ for $k\neq j$ but with the constraint that the total delay time,  $T_{tot}=4\sum_{n=1}^{n_{l}}\tau_{n}$, was the same for all experiments on a given molecule.  In Figure \ref{fig:fig9},  the FIDs (left) starting from the final echo (blue curves) along with the corresponding spectra (right) under the randomized CPMG($Y,\,Y$) pulse train in Eq. (\ref{eq:rando}) are shown for [Fig. \ref{fig:fig9}(A)] C$_{60}$, [Fig. \ref{fig:fig9}(B)] C$_{70}$, and [Fig. \ref{fig:fig9}(C)] adamantane.  The corresponding CPMG($Y,\,Y$) experiment using the averaged time delay, $\overline{\langle \tau\rangle}=\frac{1}{n_{l}}\sum_{n=1}^{n_{l}}\tau_{n}$,  with $T_{tot}=4\overline{\langle\tau\rangle} n_{l}$, is shown for comparison [red curves in Fig. \ref{fig:fig9}].  In all cases, the echo amplitudes were smaller under the randomized CPMG($Y,\,Y$) pulse train compared to the regular CPMG($Y,\,Y$) pulse train;  however, in [Fig. \ref{fig:fig9}(B)] C$_{70}$ and [Fig. \ref{fig:fig9}(C)] adamantane, this reduction in echo amplitude was mainly from a decrease in the intensity from single-quantum transitions with frequencies $\nu\neq 0$ Hz.  In fact, the intensity of the resonances near $\nu=0$ Hz was nearly the same under both the randomized and regular CPMG($Y,\,Y$) pulse trains in [Fig. \ref{fig:fig9}(B)] C$_{70}$ and [Fig. \ref{fig:fig9}(C)] adamantane.  This implies that the reduction in echo amplitude for adamantane and C$_{70}$ under the randomized CPMG($Y,\,Y$) pulse train was mainly a result of a line narrowing effect whereby the effective ``pulsed'' spin-locking field became more selective.  This is consistent with previous work\cite{Walls11} that has demonstrated improved selectivity by ``randomizing'' periodic excitation sequences.  In C$_{60}$ [Fig. \ref{fig:fig9}(A)], however, the $\approx 74\%$ reduction in echo amplitude under the randomized CPMG($Y,\,Y$) cannot be fully attributed to line narrowing.  While there was a $27.8\%$ reduction in the linewidth for C$_{60}$ under the randomized CPMG sequence, the peak height at $\nu=0$ Hz in the spectrum was also reduced by $\approx 50.5\%$.  This indicates that coherence pathways that involve changes in coherence order, such as the stimulated echo pathways, contribute to the signal at $\nu=0$ Hz.

  Within the context of coherence pathways, the fact that breaking the periodicity of the CPMG sequence led to increased line narrowing suggests that there is a correlation between coherence pathways and the corresponding transition frequencies, $\omega_{kj}^{m,m-1}$.  Such a correlation can be understood from the fact that single-quantum coherences with transition frequencies near $\nu_{kj}^{m,m-1}=0$ Hz and with phase $\pm Y$ will not change coherence order due to imperfect $\pi-$pulses, whereas for single-quantum coherences with transition frequencies $\left|\nu_{jk}^{m,m-1}\right|\gg 0$ Hz, the $\pi-$pulses would appear to be applied off-resonance.  This argument suggests that transitions with $\left|\nu_{kj}^{m,m-1}\right|\gg 0$ Hz are more likely to undergo coherence transfers due to imperfect $\pi-$pulses.  Randomizing the CPMG pulse train prevents such  coherence pathways from refocussing at the echo times, thereby resulting in a reduction in spectral intensity at transition frequencies $\left|\nu_{jk}^{m,m-1}\right|\gg 0$ Hz.  However, the pathway picture does not provide a clear explanation for the $\tau$ dependence of the line narrowing seen in Figs. \ref{fig:fig7} and \ref{fig:fig8} and would suggest that the line narrowing should increase with $n_{l}$.  The opposite trend was observed in Figs. \ref{fig:fig7}(A) and \ref{fig:fig8}(A).

    Besides the coherence pathway picture, the line narrowing under the randomized CPMG pulse train in Fig. \ref{fig:fig9} can also be understood within the context of the AHT treatment of our proposed theory given in {\bf{Section II}}.  While the randomness of the delays means that an effective Hamiltonian cannot be written over the cycle of a CPMG block as was done in Eq. (\ref{eq:avgH1}) and Eq. (\ref{eq:Havg2}), those transitions with $\nu_{kj}^{m,m-1}\approx 0$ Hz will be unaffected by the random delays under $\widehat{H}_D$ (i.e., $\omega_{kj}^{m,m-1}\tau_{k}\approx 0$ for all $\tau_{k}$), and such transitions will be effectively ``pulsed'' spin-locked during the course of the CPMG pulse train.  This explains the observed narrowing in the spectra under the randomized CPMG($Y,\,Y$) pulse train in Fig. \ref{fig:fig9}.

\begin{figure}
\includegraphics*[scale=.4]{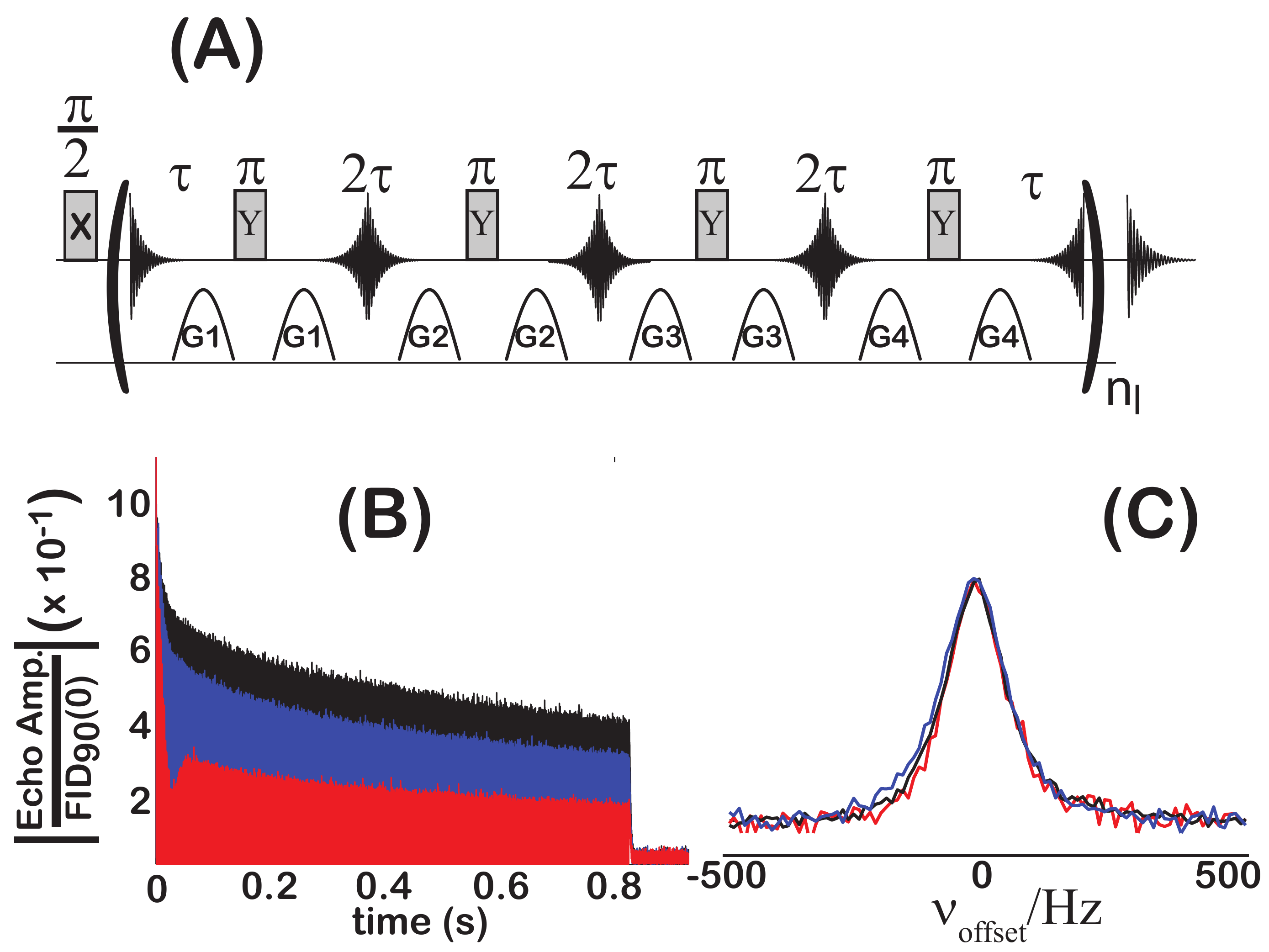}
\caption{(Color online) (A) Modified CPMG($Y,\,Y$) pulse train incorporating pulsed field gradients (PFGs) of strengths G1, G2, G3 and G4 before and after the $\pi-$pulses.  (B)  The signal from C$_{60}$ acquired during the CPMG($Y,\,Y$) sequence in Fig. \ref{fig:fig10}(A) [$\tau=1$ms, $n_{l}=100$, and N$_s$=128] using half sine shaped, 300 $\mu$s PFGs with (blue) G1 $=$ G2 $=$ G3 $=$ G4 $=$ 4.5 G/cm and (red) G1 $=$ 4.5 G/cm, G2 $=$ 3.0 G/cm, G3 $=$ 0 G/cm, and G4 $=$ 1.5 G/cm. For comparison, the CPMG($Y,\,Y$) experiment without PFGs is also shown (black). With the application of PFGs, there was a $24\%$ reduction in the echo amplitudes for (blue) the matched gradients and a $51\%$ reduction for (red) the mismatched gradients.  (C) The corresponding spectra after fourier transformation of the FID of the final echo, normalized by the corresponding echo amplitude.  The linewidth of the C$_{60}$ resonance was unaffected by the PFGs.}
\label{fig:fig10}
\end{figure}

 Since stimulated echo pathways were considered as a potential source of the long-lived echoes\cite{Franzoni08,Levstein08,Franzoni12} observed in C$_{60}$, and since we also observed a reduction in spectral intensity at $\nu\approx 0$ Hz under the randomized CPMG($Y,\,Y$) sequence in Fig. \ref{fig:fig9}(A), we performed CPMG($Y,\,Y$) pulse trains that incorporated uniaxial (along the $\widehat{z}$-direction) pulsed field gradients (PFGs) applied during the delays as shown in Fig. \ref{fig:fig10}(A) in order to control the contribution from stimulated echoes to the long-lived echoes.  The contribution from the stimulated pathway is attenuated by the application of half-sine shaped PFGs of length $t_{g}$ applied during the delays by the factor:
 \begin{eqnarray}
 \frac{1}{z_{top}-z_{bottom}}\left|\int_{z_{bottom}}^{z_{top}}e^{i\frac{2}{\pi}\gamma(G_{j}-G_{k})z't_{g}}dz'\right|
  \label{eq:atten}
  \end{eqnarray}
  where the integral in Eq. (\ref{eq:atten}) extends over the sample height ($z_{top}-z_{bottom}\approx 1$ cm in our experiment).  If the strengths of the PFGs are all equal (G1 $=$ G2 $=$ G3 $=$ G4), then the stimulated echo is not attenuated, whereas if G1 $\neq$ G2 $\neq$ G3 $\neq$ G4, the stimulated echo pathway is attenuated due to the oscillatory nature of the integrand in Eq. (\ref{eq:atten}).  The use of mismatched gradients to remove stimulated echo pathways has been previously performed in early excitation sculpting experiments\cite{Hwang95}. In Fig. \ref{fig:fig10}(B), the echo amplitudes under a CPMG$(Y,\,Y)$ experiment ($\tau=1$ ms, $n_{l}$=100) without PFGs (black), and with either matched PFGs [(blue) G1 $=$ G2 $=$ G3 $=$ G4 $=$ 4.5 G/cm, and no expected attenuation of stimulated echoes from Eq. (\ref{eq:atten})] or mismatched PFGs [(red) G1 $=$ 4.5 G/cm, G2 $=$ 3.0 G/cm, G3 $=$ 0 G/cm, and G4 $=$ 1.5 G/cm, with a minimum attenuation of the stimulated echoes of $85.4\%$ from Eq. (\ref{eq:atten})] are shown.  Compared to the echo amplitudes in the absence of PFGs, there was a $24\%$ and $51\%$ reduction in echo amplitudes for the matched and mismatched PFGs, respectively [Fig. \ref{fig:fig10}(B)].  Unlike the randomized CPMG pulse trains in Fig. \ref{fig:fig9}(A), the reduction in echo amplitude was not due to line narrowing, as can be seen from the spectra from the last echo in Fig. \ref{fig:fig10}(C), which were normalized by the echo amplitude for better comparison of the linewidths.  In this case, there was no discernible difference between the normalized spectra in Fig. \ref{fig:fig10}(C), indicating that the PFGs uniformly scale the echo amplitude intensity.  The fact that the echo amplitudes were reduced by 51$\%$ under mismatched gradients [Fig. \ref{fig:fig10}(B), red] suggests that the reduction in signal at $\nu\approx 0$ Hz under the randomized CPMG($Y,\,Y$) pulse train in Fig. \ref{fig:fig9}(A), which was $\approx 50.5\%$, comes from removing the stimulated echo pathway in C$_{60}$.  From the simulations in Fig. \ref{fig:fig2a}, which indicated that the spin dynamics within the zero-quantum/single-quantum subspace was important in establishing long-lived echoes, it should also not be surprising that attenuating the contributions of the stimulated echo pathways results in a reduction in echo amplitude.   With regard to our proposed theory in {\bf{Section II}}, matched PFGs do affect $\overline{H}_{\text{avg}}$ by making $\Delta \chi$ spatially dependent, $\Delta\chi=\frac{\phi_{2}-\phi_{1}}{2}+\delta\phi-\omega_{\text{off}}\tau-\frac{2}{\pi}\gamma G_{i}zt_{g}$.  This results in a spatial modulation of the magnitude of the pulsed ``spin-locking'' field, which attenuates the $\widehat{y}$-magnetization component of the quasiequilibrium. For example, the maximum attenuation from matched gradients should be around $50\%$ from Eq. (\ref{eq:magfin}).  For mismatched gradients, both the magnitude and direction of the the effective spin-locking field can be spatially modulated, thus reducing the quasiequilibrium magnetization even more than for the matched gradient case.

\section{Summary and Conclusions}
In conclusion, we have presented a general theory to describe the evolution of dipolar coupled spins under a CPMG($\phi_{1}\,\phi_{2}$) pulse train [Fig. \ref{fig:fig1}(B)] using imperfect $\pi-$pulses.  We have demonstrated that the form of the propagator for the CPMG($Y,\,Y$) and CPMG($X,-X$) pulse trains with imperfect $\pi-$pulses in Eq. (\ref{eq:cpmgprops}) is similar to the propagator for pulsed spin-locking of $\widehat{y}$-magnetization in dipolar solids\cite{Suwelack80,Ivanov78,Maricq85}, whereas the CPMG($X,\,X$) and CPMG($Y,-Y$) propagators can only spin-lock initial $\widehat{x}$-magnetization.  Similar to pulsed spin-locking in dipolar solids, we predicted that the CPMG($Y,\,Y$) and CPMG($X,-X$) pulse trains could generate a periodic quasiequilibrium\cite{Sakellariou98} corresponding to long-lived echoes for initial $\widehat{y}$-magnetization.  As the interpulse spacing ($2\tau$) was increased, the effective pulses under the CPMG($\phi_{1},\phi_{2}$) pulse train became less RF-like, which led to a reduction in the signal from the long-lived echoes along with a narrowing of the resulting spectrum after the application of the CPMG pulse train.  Numerical simulations along with experiments on C$_{60}$, C$_{70}$, and adamantane demonstrated this line narrowing in dipolar solids.  Randomizing the delays in the CPMG pulse train also led to a reduction in echo amplitudes, although for C$_{70}$ and adamantane, this reduction was mainly due to a line narrowing in the resulting spectra with only single-quantum coherences with transition frequencies $\nu_{kj}^{m,m-1}\approx 0$ Hz contributing to the long-lived echoes.  In C$_{60}$, however, there was both a 27.8$\%$ reduction in linewidth along with a 50.5$\%$ reduction in spectral intensity at $\nu\approx 0$ Hz.  Incorporating mismatched pulsed field gradients (PFGs) into the CPMG($Y,\,Y$) pulse train resulted in a $\approx 51\%$ reduction in intensity of the echo amplitudes in C$_{60}$, suggesting that stimulated echo pathways contribute at least $50\%$ of the signal intensity of the long-lived echoes in C$_{60}$.  With regard to the theoretical picture presented in this work, further experiments that characterize the quasiequilibrium and experiments starting with other initial conditions, such as dipolar order, are currently being performed in both solid and liquid crystalline samples.  Modifications to existing theories of pulsed spin-locking\cite{Maricq85} to characterize the rate of establishing the quasiequilibrium and its subsequent decay are also being investigated.

Finally, we will end by comparing the theory presented in this work to the Barrett\cite{Li07,Li08,Dong08} and Levstein\cite{Franzoni05,Franzoni08,Levstein08,Franzoni12} proposals.  Like both proposals, our theory predicts that the CPMG($Y,\,Y$) and CPMG($X,-X$) pulse trains can generate long-lived echoes from initial $\widehat{y}$-magnetization while the CPMG($X,\,X$) and CPMG($Y,-Y$) pulse trains cannot.  Unlike either the Barrett or Levstein proposals, however, our theory predicts a line narrowing under the CPMG pulse trains that depends on the $\pi-$pulse spacing.
{\bf{With regards to the Barrett proposal}}, inclusion of $\widehat{H}_{D}$ during the pulse does not significantly alter our theoretical results when $\tau\gg t_{p}$.  Simulations without $\widehat{H}_D$ during the pulse were almost identical to the simulations with $\widehat{H}_{D}$ included during the $\pi-$pulse (data not shown).  From the calculations in {\bf{Appendix B}}, including $\widehat{H}_{D}$ during the pulse when $\tau\gg t_{p}$ slightly attenuates the quasiequilibrium magnetization [Eq. (\ref{eq:qemmm})].  Furthermore, the propagator under a CPMG($\phi_{1},\,\phi_{2}$) pulse train was shown to be similar to the propagator under pulsed spin-locking, albeit with an ``RF-like'' Hamiltonian given by $\overline{H}_{\text{avg}}$ in Eq. (\ref{eq:Havg1}).  {\bf{With regards to the Levstein proposal}}, the [Fig. \ref{fig:fig9}] randomized CPMG and the [Fig. \ref{fig:fig10}] CPMG/PFG experiments indicated that stimulated echoes contribute to $50\%$ of the long-lived echo amplitudes in C$_{60}$.  However, it was not clear that the stimulated echo contribution to the long-lived echoes in C$_{60}$ was enhanced by the fact that T$_{1}\gg $T$_{2}$. Experiments that measure the temperature and field dependence of the long-lived echoes in order to detect a T$_{1}$ dependence on the long-lived echoes are currently in progress, along with a more careful investigation\cite{Baltisberger12} mapping out all coherence transfer pathways under the CPMG pulse trains.  Further experiments looking at quadrupolar spins under CPMG pulse trains under static conditions are also being investigated.

\section{Acknowledgments}
 This material is based upon work supported by the National Science Foundation under CHE - 1056846, the Camille and Henry Dreyfus Foundation, and the University of Miami.

\appendix
\section{Details of the Average Hamiltonian Theory Calculations in Section II}
Starting from Eq. (\ref{eq:impii}) and transforming into an interaction frame given by $\widehat{V}(t)=e^{-i\omega_{RF}t(\widehat{I}_{X}\cos(\phi)+\widehat{I}_{Y}\sin(\phi))}$, the propagator for an imperfect $\pi-$pulse, $\widehat{R}_{\phi}(\pi)$, can be rewritten as [with $\omega_{RF}t_{p}=\pi$ and $\widehat{V}(t_{p})=\widehat{P}_{\frac{\pi}{2},\phi}(\pi)$]:
\begin{eqnarray}
\widehat{R}_{\phi}(\pi)&=&\widehat{P}_{\frac{\pi}{2},\phi}(\pi)Te^{-i\left[\int^{t_{p}}_{0}\text{d}t'\delta\omega_{RF}\left(\widehat{I}_{X}\cos(\phi)+\widehat{I}_{Y}\sin(\phi)\right)+\omega_{\text{off}}\widehat{V}^{\dagger}(t')\widehat{I}_{Z}\widehat{V}(t')\right]}\nonumber\\
&=&\widehat{P}_{\frac{\pi}{2},\phi}(\pi)Te^{-i\left[\int^{t_{p}}_{0}\text{d}t'\delta\omega_{RF}\left(\widehat{I}_{X}\cos(\phi)+\widehat{I}_{Y}\sin(\phi)\right)+\omega_{\text{off}}\left(\widehat{I}_{Z}\cos(\omega_{RF}t')+\sin(\omega_{RF}t')(\widehat{I}_{Y}\cos(\phi)-\widehat{I}_{X}\sin(\phi))\right)\right]}\nonumber\\
&\equiv&\widehat{P}_{\frac{\pi}{2},\phi}(\pi)Te^{-\frac{i}{\hbar}\int^{t_{p}}_{0}\widetilde{H}(t')\text{d}t'}
\end{eqnarray}
where $T$ is the Dyson time-ordering operator.  Average Hamiltonian theory\cite{Haeberlen68} can be used to approximate $Te^{-\frac{i}{\hbar}\int^{t_p}_{0}\text{d}t'\widetilde{H}(t')}\approx e^{-i\frac{t_{p}}{\hbar}\overline{H}_{\text{avg}}}$, where $\overline{H}_{\text{avg},\phi}$ is the average Hamiltonian over the time $t_{p}$, which can be written as $\overline{H}_{\text{avg},\phi}=\sum_{n=1}^{\infty}\overline{H}^{(n)}_{\text{avg},\phi}$ with $\overline{H}_{\text{avg},\phi}^{(n)}$ being the $n^{th}$-order average Hamiltonian.  An expression for $\overline{H}_{\text{avg},\phi}^{(1)}$ is given by:
\begin{eqnarray}
\frac{\overline{H}_{\text{avg},\phi}^{(1)}}{\hbar}&=&\frac{1}{\hbar t_{p}}\int^{t_{p}}_{0}\text{d}t'\widetilde{H}(t')=\delta\omega_{RF}\left(\widehat{I}_{X}\cos(\phi)+\widehat{I}_{Y}\sin(\phi)\right)\nonumber\\
&+&\frac{\omega_{\text{off}}}{t_{p}}\left(\widehat{I}_{Z}\int^{t_{p}}_{0}\cos(\omega_{RF}t')\text{d}t'+(\widehat{I}_{Y}\cos(\phi)-\widehat{I}_{X}\sin(\phi))\int^{t_{p}}_{0}\text{d}t'\sin(\omega_{RF}t')\right)\nonumber\\
&=&\delta\omega_{RF}\left(\widehat{I}_{X}\cos(\phi)+\widehat{I}_{Y}\sin(\phi)\right)-\frac{\omega_{\text{off}}}{\omega_{RF}t_{p}}\cos(\omega_{RF}t)|^{t=t_{p}}_{t=0}\left(\widehat{I}_{Y}\cos(\phi)-\widehat{I}_{X}\sin(\phi)\right)\nonumber\\
&=&\delta\omega_{RF}\left(\widehat{I}_{X}\cos(\phi)+\widehat{I}_{Y}\sin(\phi)\right)+\frac{2\omega_{\text{off}}}{\pi}\left(\widehat{I}_{Y}\cos(\phi)-\widehat{I}_{X}\sin(\phi)\right)\nonumber\\
&=&\frac{1}{2}\left(\widehat{I}_{+}e^{-i\phi}\left(\delta\omega_{RF}-i\frac{2\omega_{\text{off}}}{\pi}\right)+\widehat{I}_{-}e^{i\phi}\left(\delta\omega_{RF}+i\frac{2\omega_{\text{off}}}{\pi}\right)\right)\nonumber\\
&=&\frac{\delta\omega}{2}\left(\widehat{I}_{+}e^{-i(\phi-\delta\phi)}+\widehat{I}_{-}e^{i(\phi+\delta\phi)}\right)
\label{eq:impir}\end{eqnarray}
which is the average Hamiltonian given in the first line of Eq. (\ref{eq:impi}).  Note that the subscript $\phi$ denotes the phase of the $\pi$ pulse.

Similarly, $\overline{H}_{\text{avg},\phi}^{(2)}$ is given by:
\begin{eqnarray}
\frac{\overline{H}_{\text{avg},\phi}^{(2)}}{\hbar}&=&\frac{1}{2i\hbar^{2}t_{p}}\int^{t_{p}}_{0}\text{d}t'\int^{t'}_{0}\text{d}t''\left[\widetilde{H}(t'),\widetilde{H}(t'')\right]=-\frac{2\delta\omega_{RF}\omega_{\text{off}}}{\pi\omega_{RF}}\left(\widehat{I}_{Y}\cos(\phi)-\widehat{I}_{X}\sin(\phi)\right)\nonumber\\
&+&\frac{\omega^{2}_{\text{off}}}{2\omega_{RF}}\left(\widehat{I}_{X}\cos(\phi)+\widehat{I}_{Y}\sin(\phi)\right)
\label{eq:apH2}
\end{eqnarray}
As can be seen from Eq. (\ref{eq:apH2}), the terms in $\overline{H}^{(2)}_{\text{avg},\phi}$ are scaled by factors of $\frac{\delta\omega_{RF}}{\omega_{RF}}$ and/or $\frac{\omega_{\text{off}}}{\omega_{RF}}$ relative to the terms in $\overline{H}^{(1)}_{\text{avg},\phi}$.  In general, the terms in $\overline{H}^{(n)}_{\text{avg},\phi}$ will be scaled by factors of $\left(\frac{\delta\omega_{RF}}{\omega_{RF}}\right)^{p}\left(\frac{\omega_{\text{off}}}{\omega_{RF}}\right)^{n-1-p}$ for $p\in [0, n-1]$ relative to the terms in $\overline{H}_{\text{avg},\phi}^{(1)}$.  Therefore, if $\left|\frac{\delta\omega_{RF}}{\omega_{RF}}\right|\ll 1$ and $\left|\frac{\omega_{\text{off}}}{\omega_{RF}}\right|\ll 1$, then the approximation $\overline{H}_{\text{avg},\phi}\approx \overline{H}_{\text{avg},\phi}^{(1)}$ in Eq. (\ref{eq:impi}) is justified.

For the CPMG($\phi_{1},\,\phi_{2}$) pulse block, the propagator is given by:
\begin{eqnarray}
\widehat{U}_{\text{CPMG}_{\phi_{1},\,\phi_{2}}}(4\tau)&\approx& \widehat{U}_{f}(\tau)\widehat{P}_{\frac{\pi}{2},\phi_{2}}(\pi)e^{-i\frac{t_{p}}{\hbar}\overline{H}_{\text{avg},\phi_{2}}}\widehat{U}_{f}(2\tau)\widehat{P}_{\frac{\pi}{2},\phi_{1}}(\pi)e^{-i\frac{t_{p}}{\hbar}\overline{H}_{\text{avg},\phi_{1}}}\widehat{U}_{f}(\tau)
\label{eq:ucpmg}
\end{eqnarray}
Rewriting $\widehat{U}_{f}$ in the interaction frame defined by $\widehat{V}(t)=e^{-i\frac{t}{\hbar}(\widehat{H}_{D}+\hbar\omega_{\text{off}}\widehat{I}_{Z})}=\widehat{U}_{D}(\tau)\widehat{P}_{0,0}(\omega_{\text{off}}\tau)$ as
\begin{eqnarray}
\widehat{U}_{f}(\tau)&=&\widehat{V}(\tau)Te^{-\frac{i}{\hbar}\int^{\tau}_{0}\text{d}t'\widehat{V}^{\dagger}(t')\widehat{H}_{cs}\widehat{V}(t')}\nonumber\\
&\approx&\widehat{U}_{D}(\tau)\widehat{P}_{0,0}(\omega_{\text{off}}\tau)e^{-\frac{i\tau}{\hbar}\overline{H}^{(1)}_{cs,\text{avg}}(\tau)}
\end{eqnarray}
where
\begin{eqnarray}
\frac{\overline{H}^{(1)}_{cs,\text{avg}}(\tau)}{\hbar}&=&\frac{1}{\hbar\tau}\int^{\tau}_{0}\text{d}t'\widehat{V}^{\dagger}(t')\widehat{H}_{cs}\widehat{V}(t')=\sum_{k,m}\left(H_{cs}\right)_{k,k}^{m,m}|\epsilon_{k},m\rangle\langle\epsilon_{k},m|\nonumber\\
&+&\sum_{(k<j),m}\text{sinc}\left(\frac{\omega_{jk}^{m,m}\tau}{2}\right)\left(e^{\frac{i\omega_{jk}^{m,m}\tau}{2}}\left(H_{cs}\right)_{j,k}^{m,m}|\epsilon_{j},m\rangle\langle\epsilon_{k},m|+e^{-i\frac{\omega_{jk}^{m,m}\tau}{2}}\left(H_{cs}\right)^{m,m}_{j,k}|\epsilon_{k},m\rangle\langle\epsilon_{j},m|\right)\nonumber\\
\end{eqnarray}
Using the fact that $\widehat{P}_{\frac{\pi}{2},\phi_{2}}(\pi)\widehat{P}_{\frac{\pi}{2},\phi_{1}}(\pi)=\widehat{P}_{0,0}(2(\phi_{2}-\phi_{1}))$, both the $\pi-$pulses, $\widehat{P}_{\frac{\pi}{2},\phi_{1}}(\pi)$ and $\widehat{P}_{\frac{\pi}{2},\phi_{2}}(\pi)$, and the dipolar evolution propagator, $\widehat{U}_{D}(\tau)$, can be propagated through in Eq. (\ref{eq:ucpmg}) to give:
\begin{eqnarray}
\widehat{U}_{\text{CPMG}_{\phi_{1},\,\phi_{2}}}(4\tau)&\approx&\widehat{P}_{0,0}(2(\phi_{2}-\phi_{1}))\widehat{U}_{D}(4\tau)e^{-\frac{i\tau}{\hbar}\widehat{U}^{\dagger}_{D}(3\tau)\overline{H}_{cs,\text{avg}}^{(1)}(\tau)\widehat{U}_{D}(3\tau)}\nonumber\\
&\times&e^{-\frac{it_{p}}{\hbar}\widehat{U}_{D}^{\dagger}(3\tau)\widehat{P}_{0,0}(\omega_{\text{off}}\tau)\widehat{P}^{\dagger}_{\frac{\pi}{2},\phi_{1}}(\pi)\overline{H}_{\text{avg},\phi_{2}}\widehat{P}_{\frac{\pi}{2},\phi_{1}}(\pi)\widehat{P}^{\dagger}_{0,0}(\omega_{\text{off}}\tau)\widehat{U}_{D}(3\tau)}\nonumber\\
&\times&e^{\frac{2i\tau}{\hbar}\widehat{U}^{\dagger}_{D}(\tau)\overline{H}_{cs,\text{avg}}^{(1)}(2\tau)\widehat{U}_{D}(\tau)}\times e^{-\frac{it_{p}}{\hbar}\widehat{U}^{\dagger}_{D}(\tau)\widehat{P}^{\dagger}_{0,0}(\omega_{\text{off}}\tau)\overline{H}_{\text{avg},\phi_{1}}^{(1)}\widehat{P}_{0,0}(\omega_{\text{off}}\tau)\widehat{U}_{D}(\tau)}\nonumber\\
&\times&e^{-\frac{i\tau}{\hbar}\overline{H}_{cs,\text{avg}}^{(1)}(\tau)}\approx\widehat{P}_{0,0}(2(\phi_{2}-\phi_{1}))\widehat{U}_{D}(2\tau)e^{-\frac{i2t_{p}}{\hbar}\overline{H}_{\text{avg}}}\widehat{U}_{D}(2\tau)
\end{eqnarray}
where $\overline{H}_{\text{avg}}$ is the effective average Hamiltonian in the {\emph{dipolar interaction frame}}. The first-order contribution to $\overline{H}_{\text{avg}}$ is given by:
\begin{eqnarray}
\frac{\overline{H}^{(1)}_{\text{avg}}}{\hbar}&=&\frac{\tau}{2\hbar t_{p}}\left(\widehat{U}_{D}^{\dagger}(\tau)\overline{H}_{cs,\text{avg}}^{(1)}(\tau)\widehat{U}_{D}(\tau)+\widehat{U}_{D}(2\tau)\overline{H}_{cs,\text{avg}}^{(1)}(\tau)\widehat{U}^{\dagger}_{D}(2\tau)\right)\nonumber\\
&-&\frac{\tau}{\hbar t_{p}}\widehat{U}_{D}(\tau)\overline{H}_{cs,\text{avg}}(2\tau)\widehat{U}^{\dagger}_{D}(\tau)+\frac{1}{2\hbar}\widehat{U}_{D}(\tau)\widehat{P}_{0,0}^{\dagger}(\omega_{\text{off}}\tau)\overline{H}_{\text{avg},\phi_{1}}^{(1)}\widehat{P}_{0,0}(\omega_{\text{off}}\tau)\widehat{U}_{D}^{\dagger}(\tau)\nonumber\\
&+&\frac{1}{2\hbar}\widehat{U}_{D}^{\dagger}(\tau)\widehat{P}_{0,0}(\omega_{\text{off}}\tau)\widehat{P}_{\frac{\pi}{2},\phi_{1}}^{\dagger}(\pi)\overline{H}_{\text{avg},\phi_{2}}\widehat{P}^{\dagger}_{\frac{\pi}{2},\phi_{1}}(\pi)\widehat{P}^{\dagger}_{0,0}(\omega_{\text{off}}\tau)\widehat{U}_{D}(\tau)\nonumber\\
&=&\frac{2\tau}{t_{p}}\sum_{(k<j),m}(\text{sinc}\left[2\omega_{jk}^{m,m}\tau\right]-\text{sinc}\left[\omega_{jk}^{m,m}\tau\right])\left(\left(H_{cs}\right)^{m,m}_{j,k}|\epsilon_{j},m\rangle\langle\epsilon_{k},m|+\left(H_{cs}\right)^{m,m}_{k,j}|\epsilon_{k},m\rangle\langle\epsilon_{j},m|\right)\nonumber\\
&+&\frac{\delta\omega}{4}\left(\widehat{U}^{\dagger}_{D}(\tau)\widehat{I}_{+}\widehat{U}_{D}(\tau)e^{i\Delta\chi}+\widehat{U}_{D}(\tau)\widehat{I}_{+}\widehat{U}_{D}^{\dagger}(\tau)e^{-i\Delta\chi}\right)e^{i\Psi_{1}}\nonumber\\
&+&\frac{\delta\omega}{4}\left(\widehat{U}^{\dagger}_{D}(\tau)\widehat{I}_{-}\widehat{U}_{D}(\tau)e^{-i\Delta\chi}+\widehat{U}_{D}(\tau)\widehat{I}_{-}\widehat{U}_{D}^{\dagger}(\tau)e^{i\Delta\chi}\right)e^{-i\Psi_{1}}
\label{eq:rederive}
\end{eqnarray}
Eq. (\ref{eq:rederive}) is the same expression for $\overline{H}_{\text{avg}}^{(1)}$ as given in Eq. (\ref{eq:avgH1}).
\section{Including $\widehat{H}_{cs}$ and $\widehat{H}_{D}$ during the $\pi-$pulses in the CPMG($\phi_{1},\,\phi_{2}$) pulse train}
In the following, the lowest order contributions of both $\widehat{H}_{cs}$ and $\widehat{H}_{D}$ during the $\pi-$pulses in the CPMG($\phi_{1},\,\phi_{2}$) pulse trains is presented.  However, including $\widehat{H}_{cs}$ and $\widehat{H}_{D}$ during the $\pi-$pulses results in a slight modification of the theoretical framework presented in {\bf{Section II}}.
\subsection{Contributions of $\widehat{H}_{cs}$ in the CPMG($\phi_{1},\,\phi_{2}$) pulse block}
The contribution of $\widehat{H}_{cs}$ during a $\pi-$pulse of phase $\phi$ to $\overline{H}_{\text{avg}}^{(1)}$ in Eq. (\ref{eq:impir}) is given by:
\begin{eqnarray}
\frac{\overline{H}^{(1)}_{\text{avg},cs}}{\hbar}=\sum_{j=1}^{N}\frac{2\omega_{cs}^{j}}{\pi}\left(\widehat{I}_{Y,j}\cos(\phi)-\widehat{I}_{X,j}\sin(\phi)\right)
\label{eq:hcso}
\end{eqnarray}
which is similar in form to the contribution from a resonance offset given in Eq. (\ref{eq:impir}).

Using Eq. (\ref{eq:hcso}), the contribution from $\widehat{H}_{cs}$ during the $\pi-$pulses in the CPMG($\phi_{1},\,\phi_{2}$) pulse block can be accounted for by making the following changes to $\overline{H}_{\text{avg}}$ in Eq. (\ref{eq:rederive}):   $\delta\omega\rightarrow \delta\omega_{j}=\sqrt{\delta\omega_{RF}^{2}+\frac{4}{\pi^{2}}(\omega_{\text{off}}+\omega_{cs}^{j})^{2}}$, $e^{-i\delta\phi}\rightarrow e^{-i\delta\phi_{j}}=\frac{\delta\omega_{RF}-\frac{2i}{\pi}(\omega_{\text{off}}+\omega_{cs}^{j})}{\delta\omega_{j}}$,  $\Delta\chi\rightarrow \Delta\chi_{j}=\frac{\phi_{2}-\phi_{1}}{2}+\delta\phi_{j}-\omega_{\text{off}}\tau$, and $\delta \omega\widehat{I}_{\pm}e^{i\Delta\chi}\rightarrow \sum_{j}\delta\omega_{j}\widehat{I}_{\pm,j}e^{i\Delta\chi_{j}}$.

Including $\overline{H}^{(1)}_{\text{avg},cs}$ in Eq. (\ref{eq:hcso}) and also including the contributions of $\widehat{H}_{cs}$ during the $\tau$ delays (which was neglected for simplicity in $\overline{H}_{\text{avg}}^{(1)}$ in Eq. (\ref{eq:Havg1})), the quasiequilibrium magnetization under the CPMG($\phi_{1},\,\phi_{2}$) pulse train, $\langle \widehat{I}_{Y}\rangle_{QE}$ in Eq. (\ref{eq:magfin}) can be written as:
\begin{eqnarray}
\frac{\langle \widehat{I}_{Y}\rangle_{QE}}{\langle \widehat{I}_{Y}(0)\rangle}&\approx&\frac{\sin^{2}(\Psi_{1})\left[\sum_{k}\delta\omega_{k}\cos(\Delta\chi_{k})\text{FID}_{D}^{k}(\tau)\right]^{2}}{B_{p}(\tau)+B_{\text{delay}}(\tau)+\left(\frac{2\tau}{t_{p}}\right)^{2}\left(\omega_{loc}\right)^{2}}
\label{eq:qemmcs}
\end{eqnarray}
where
\begin{eqnarray}
\text{FID}_{D}^{k}(\tau)&=&\frac{\text{Trace}\left[\widehat{I}_{X,k}\widehat{U}_{D}(\tau)\widehat{I}_{X}\widehat{U}_{D}^{\dagger}(\tau)\right]}{\text{Trace}\left[\left(\widehat{I}_{X}\right)^{2}\right]}\nonumber\\
\text{FID}_{D}^{kj}(\tau)&=&\frac{\text{Trace}\left[\widehat{I}_{X,k}\widehat{U}_{D}(\tau)\widehat{I}_{X,j}\widehat{U}_{D}^{\dagger}(\tau)\right]}{\text{Trace}\left[\left(\widehat{I}_{X}\right)^{2}\right]}\nonumber\\
B_{p}(\tau)&=&\frac{\sum_{k}\delta\omega_{k}^{2}\left(1+\cos(2\Delta\chi_{k})\text{FID}_{D}^{kk}(2\tau)\right)+\sum_{k<j}\delta\omega_{k}\delta\omega_{j}\cos(\Delta\chi_{k}+\Delta\chi_{j})\text{FID}_{D}^{kj}(2\tau)}{2}\nonumber\\
B_{\text{delay}}(\tau)&=&\sum_{(k,m)<(l,m)}\left|\frac{2\tau}{t_{p}}\left(\widehat{H}_{cs}\right)_{k,l}\left(\text{sinc}[2\omega_{kl}^{m,m}\tau]-\text{sinc}[\omega_{kl}^{m,m}\tau]\right)\right|^{2}\frac{\text{Trace}\left[|\epsilon_{k},m\rangle\langle\epsilon_{k},m|+|\epsilon_{l},m\rangle\langle\epsilon_{l},m|\right]}{\text{Trace}\left[\left(\widehat{I}_{Y}\right)^{2}\right]}
\label{eq:hcsb}
\end{eqnarray}
In Eq. (\ref{eq:qemmcs}), $B_{p}(\tau)$ is due to the contribution of $\widehat{H}_{cs}$ during the $\pi-$pulse, and $B_{\text{delay}}(\tau)$ is due to the contribution of $\widehat{H}_{cs}$ during the $\tau$ delays.
 \subsection{Contributions of $\widehat{H}_{D}$ in the CPMG($\phi_{1},\,\phi_{2}$) pulse block}
The contribution of $\widehat{H}_{D}$ during a $\pi-$pulse of phase $\phi$ to $\overline{H}_{\text{avg}}^{(1)}$ in Eq. (\ref{eq:impir}) is given by
\begin{eqnarray}
\overline{H}^{(1)}_{\text{avg},D}&=&\frac{1}{4}\widehat{H}_{D}-\frac{3}{8}\sum_{j<k}\hbar\omega_{D}^{jk}\left(\widehat{I}_{+,j}\widehat{I}_{+,k}e^{-2i\phi}+\widehat{I}_{-,j}\widehat{I}_{-,k}e^{2i\phi}\right)
\label{eq:hdo}
\end{eqnarray}

Using Eq. (\ref{eq:hdo}), the contribution of $\widehat{H}_{D}$ during the $\pi-$pulses in the CPMG($\phi_{1},\,\phi_{2}$) pulse block can be accounted for by adding the following to $\overline{H}_{\text{avg}}$ in Eq. (\ref{eq:rederive}):
{\small{
\begin{eqnarray}
\overline{H}_{\text{avg},D}&=&\frac{\widehat{H}_{D}}{4}-\sum_{j<k}\frac{3\hbar\omega_{D}^{jk}}{16}\left(\widehat{U}_{D}(\tau)\widehat{I}_{+,j}\widehat{I}_{+,k}\widehat{U}^{\dagger}_{D}(\tau)e^{i(\phi_{1}-\phi_{2}+2\omega_{\text{off}}\tau)}+\widehat{U}^{\dagger}_{D}(\tau)\widehat{I}_{+,j}\widehat{I}_{+,k}\widehat{U}_{D}(\tau)e^{-i(\phi_{1}-\phi_{2}+2\omega_{\text{off}}\tau)}\right)e^{2i\Psi_1}\nonumber\\
&-&\sum_{j<k}\frac{3\hbar\omega_{D}^{jk}}{16}\left(\widehat{U}_{D}(\tau)\widehat{I}_{-,j}\widehat{I}_{-,k}\widehat{U}^{\dagger}_{D}(\tau)e^{-i(\phi_{1}-\phi_{2}+2\omega_{\text{off}}\tau)}+\widehat{U}^{\dagger}_{D}(\tau)\widehat{I}_{-,j}\widehat{I}_{-,k}\widehat{U}_{D}(\tau)e^{i(\phi_{1}-\phi_{2}+2\omega_{\text{off}}\tau)}\right)e^{-2i\Psi_1}\nonumber\\
\label{eq:hd}
\end{eqnarray}
}}
 Including $\overline{H}_{\text{avg},D}$ in $\overline{H}^{(1)}_{\text{avg}}$ [Eq. (\ref{eq:rederive})], the quasiequilibrium magnetization under the CPMG($\phi_{1},\,\phi_{2}$) pulse train, $\langle \widehat{I}_{Y}\rangle_{QE}$ in Eq. (\ref{eq:qemmcs}), is attenuated and is given by:
\begin{eqnarray}
\frac{\langle \widehat{I}_{Y}\rangle_{QE}}{\langle \widehat{I}_{Y}(0)\rangle}&\approx&\frac{\sin^{2}(\Psi_{1})\left[\sum_{k}\delta\omega_{k}\cos(\Delta\chi_{k})\text{FID}_{D}^{k}(\tau)\right]^{2}}{B_{p}(\tau)+B_{\text{delay}}(\tau)+\left[\left(\frac{2\tau+t_p}{t_{p}}\right)^{2}+\frac{3}{64}\right]\left(\omega_{loc}\right)^{2}+\frac{9}{128}A_{2Q}(2\tau)}
\label{eq:qemmm}
\end{eqnarray}
where
\begin{eqnarray}
A_{2Q}(2\tau)&=&\text{Real}\left[\sum_{j<k}\left(\omega_{D}^{jk}\right)^{2}\text{Tr}\left[\widehat{U}_{D}(2\tau)\widehat{I}_{+,j}\widehat{I}_{+,k}\widehat{U}^{\dagger}_{D}(2\tau)\widehat{I}_{-,j}\widehat{I}_{-,k}\right]e^{2i(\phi_{1}-\phi_{2}-2\omega_{\text{off}}\tau)}\right]
\end{eqnarray}
represents the self correlation of double-quantum dipolar order under $\widehat{U}_{D}(\tau)$.  When $\tau\gg t_{p}$, this correlation is expected to be small ($A_{2Q}(\tau)\rightarrow 0$, and $\left[\left(\frac{2\tau+t_{p}}{t_{p}}\right)^{2}+\frac{3}{64}\right]\rightarrow\left(\frac{2\tau}{t_{p}}\right)^{2}$, so that the quasiequilibrium magnetization in Eq. (\ref{eq:qemmm}) reduces to Eq. (\ref{eq:qemmcs}) where $\widehat{H}_{D}$ was neglected during the $\pi$-pulses.
\subsection{$\overline{H}_{\text{avg}}$ in the limit as $\tau\rightarrow 0$}
In writing the quasiequilibrium under the CPMG($\phi_{1},\,\phi_{2}$) pulse train in Eq. (\ref{eq:quasi}), it was assumed that the only constant of motion was the effective Hamiltonian under the CPMG($\phi_{1},\,\phi_{2}$), $\widetilde{H}_{\text{eff}}$.  Under strong RF irradiation, and in the limit that $\tau\rightarrow 0$, the effective Hamiltonian (to lowest order and neglecting the effects of $\widehat{H}_{cs}$ during the $\pi$-pulses) is given by combining Eq. (\ref{eq:Havg1}) and Eq. (\ref{eq:hd}) in the limit that $\tau\rightarrow 0$:
\begin{eqnarray}
\lim_{\tau\rightarrow 0}\frac{\widetilde{H}_{\text{eff}}}{\hbar}&=&\delta\omega\cos\left(\frac{\phi_{2}-\phi_{1}}{2}+\delta\phi\right)\left(\widehat{I}_{X}\cos(\Psi_{1})-\widehat{I}_{Y}\sin(\Psi_{1})\right)+\frac{\widehat{H}_{D}}{4\hbar}\nonumber\\
&-&\frac{3}{8}\cos(\phi_{2}-\phi_{1})\sum_{j<k}\omega_{D}^{jk}\left(\widehat{I}_{+,j}\widehat{I}_{k,+}e^{2i\Psi_{1}}+\widehat{I}_{-,j}\widehat{I}_{-,k}e^{-2i\Psi_{1}}\right)
\label{eq:notau}
\end{eqnarray}
For a CPMG($Y,\,Y$) sequence, Eq. (\ref{eq:notau}) becomes \begin{eqnarray}
\frac{\widetilde{H}_{\text{eff}}}{\hbar}&=&\delta\omega\cos(\delta\phi)\widehat{I}_{Y}+\frac{\widehat{H}_{D}}{4\hbar}+\frac{3}{4}\sum_{j<k}\omega_{D}^{jk}(\widehat{I}_{X,j}\widehat{I}_{X,k}-\widehat{I}_{Y,j}\widehat{I}_{Y,k})\nonumber\\
&=&\delta\omega\cos(\delta\phi)\widehat{I}_{Y}-\frac{1}{2}\sum_{j<k}\omega_{D}^{jk}\left(2\widehat{I}_{Y,j}\widehat{I}_{Y,k}-\widehat{I}_{X,j}\widehat{I}_{X,k}-\widehat{I}_{Z,j}\widehat{I}_{Z,k}\right)\nonumber\\
&=&\delta\omega\cos(\delta\phi)\widehat{I}_{Y}-\frac{1}{2\hbar}\widehat{H}_{D,YY}
\label{eq:0tauYY}
\end{eqnarray}
where $\widehat{H}_{D,YY}$ is the dipolar Hamiltonian that is quantized along the $\widehat{y}-$direction.  In this case, $[\widehat{I}_{Y},\,\widehat{H}_{D,YY}]=0$, and so $\widetilde{H}_{\text{eff}}$ in Eq. (\ref{eq:0tauYY}) has two constants of motion, $\widehat{I}_{Y}$ and $\widehat{H}_{D,YY}$.  In principle, two initial ``temperatures'' can be associated with these constants of motion, which can be calculated using the Provotorov equations\cite{Order}.

For the CPMG($X,-X$) sequence, Eq. (\ref{eq:notau}) becomes \begin{eqnarray}
\frac{\widetilde{H}_{\text{eff}}}{\hbar}&=&\delta\omega\sin(\delta\phi)\widehat{I}_{Y}+\frac{\widehat{H}_{D}}{4\hbar}-\frac{3}{4}\sum_{j<k}\omega_{D}^{jk}(\widehat{I}_{X,j}\widehat{I}_{X,k}-\widehat{I}_{Y,j}\widehat{I}_{Y,k})\nonumber\\
&=&\delta\omega\cos(\delta\phi)\widehat{I}_{Y}-\frac{1}{2}\sum_{j<k}\omega_{D}^{jk}\left(2\widehat{I}_{X,j}\widehat{I}_{X,k}-\widehat{I}_{Y,j}\widehat{I}_{Y,k}-\widehat{I}_{Z,j}\widehat{I}_{Z,k}\right)\nonumber\\
&=&\delta\omega\cos(\delta\phi)\widehat{I}_{Y}-\frac{1}{2\hbar}\widehat{H}_{D,XX}
\label{eq:0tauXX}
\end{eqnarray}
where $\widehat{H}_{D,XX}$ is the dipolar Hamiltonian that is quantized along the $\widehat{x}-$direction.  Since $[\widehat{I}_{Y},\,\widehat{H}_{D,XX}]\neq 0$, $\widetilde{H}_{\text{eff}}$ in Eq. (\ref{eq:0tauXX}) is the only constant of motion, and so the quasiequilibrium is determined by $\widetilde{H}_{\text{eff}}$ in Eq. (\ref{eq:0tauXX}).  It is thus expected that the quasiequilibrium magnetization under the CPMG($Y,\,Y$) pulse train in the limit of $\tau\rightarrow 0$ will be larger than that under the CPMG($X,-X$), which will only exist for nonzero chemical shifts and/or offsets, $\omega_{\text{off}}\neq 0$.

\end{document}